\documentclass[aps,showpacs,preprintnumbers,amsmath, amssymb,nofootinbib]{revtex4}

\usepackage{float}
\usepackage{graphics,epsfig}
\usepackage{bm}
\usepackage{slashed}
\usepackage{graphicx}
\usepackage{amsmath}
\usepackage{amsfonts}
\usepackage{amssymb}
\usepackage{color}%
\usepackage{dcolumn}
\providecommand{\U}[1]{\protect\rule{.1in}{.1in}}

\begin{document}
\baselineskip=0.6 cm\title{Dynamically generated gap from
holography in the charged black brane with hyperscaling violation}
\author{Xiao-Mei Kuang}
\email{xmeikuang@gmail.com} \affiliation{Department of
Physics, National Technical University of Athens, GR-15780 Athens,
Greece}
\author{Eleftherios Papantonopoulos}
\email{lpapa@central.ntua.gr} \affiliation{Department of Physics,
National Technical University of Athens, GR-15780 Athens,
Greece,\\
and CERN - Theory Division, CH-1211 Geneva 23, Switzerland.}
\author{Bin Wang}
\email{wang_b@sjtu.edu.cn} \affiliation{IFSA Collaborative Innovation Center, Department of Physics and
Astronomy, Shanghai Jiao Tong University, Shanghai 200240, China.}
\author{Jian-Pin Wu}
\email{jianpinwu@gmail.com} \affiliation{Department of Physics, School of Mathematics and Physics,
Bohai University, Jinzhou, 121013, China,\\
and State Key Laboratory of Theoretical Physics, Institute of Theoretical Physics, Chinese Academy of Sciences, Beijing 100190, China.}

\date{\today}
\vspace*{0.2cm}
\begin{abstract}
\baselineskip=0.6 cm
\begin{center}
{\bf Abstract}
\end{center}
We holographically investigate the effects  of a
dipole coupling between a fermion field and a
$U(1)$  gauge field on the dual fermionic sector
in the charged gravity bulk with hyperscaling
violation.  We analytically study the features of
the ultraviolet and infrared Green's functions of
the dual fermionic system and we show that as the
dipole coupling and the hyperscaling violation
exponent are varied, the fluid possess Fermi,
marginal Fermi, non-Fermi liquid phases and also
an additional Mott insulating phase. We find that
the increase of the  hyperscaling violation
exponent which effectively reduces the
dimensionality of the system makes it harder for
the Mott gap to be formed. We also show that the
observed duality between zeros and poles in the
presence of a dipole moment coupling still
persists in theories with hyperscaling violation.

\end{abstract}

\maketitle
\newpage
\vspace*{0.2cm}
\section{Introduction}
The gauge/gravity duality is a powerful principle
which describes many physical systems
holographically. This holographic description
relates  strongly-coupled quantum field theories
with their classical gravitational counterparts
which live in one higher dimension
\cite{Maldacena:1997re,Gubser:2002,Witten:1998}.
A remarkable implementation of this holographic
principle was realized in investigating strongly
coupled systems in condensed matter physics, the
so called AdS/CMT correspondence, which made
possible the better understanding of numerous
exotic but very important features of electronic
materials, including the high temperature
superconductors and the heavy fermion systems.

To describe these systems holographycally,
fermion fields coupled to  $U(1)$ gauge field
have to be introduced in the gravity sector. The
fermions were treated as a probe and their
backreaction  on the background geometry were
ignored
\cite{Lee:2008,Liu:2009,Cubrovic:2009ye,Faulkner:2009}.
 The
properties of the spectral function showed that a
Fermi surface usually emerges, the low energy
excitations of it can exhibit Fermi liquid,
marginal Fermi liquid, or non-Fermi liquid
behaviors. To describe better  various phases of
a metallic state at low temperatures, a dipole
coupling to massless charged fermions  was
introduced \cite{Edalati:2010a,Edalati:2010b}.
The presence of the dipole moment introduced a
scale in the system and this enabled in the dual
field theory to model a Mott insulating phase,
generate dynamically a gap  and spectral weight
transfer. This proposal has triggered  further
interest in the study of the dipole coupling
effect on the holographic fermionic systems
\cite{Guarrera:2011my,Wu:2011a,Kuang:2012,Fang:2013,Wu:2011b,Li:2011,Ling:2014}.
Then it was found in  \cite{Alsup:2014uca}  that
there exists a duality between zeroes and poles
in holographic systems with massless fermions and
a dipole coupling, which  was also observed in
\cite{Vanacore:2014hka}.

In many condensed matter materials it was
observed that at criticality their scaling
properties go beyond  the standard Lorentz
scaling. Then there is a need to describe
holographically these systems
 with  anisotropic (Lifshitz) scaling
characterized by the dynamic critical exponent $z > 1$
\cite{Kachru:2008}, or even with hyperscaling violation
characterized by a non-zero hyperscaling violation exponent
$\theta$  \cite{Gouteraux:2011,Huijse:2012,Dong:2012}.

To  formulate the duality principle to describe
holographically of such systems, it was proposed
\cite{Charmousis:2010} that their gravity duals
should have a metric of the form
\begin{equation}\label{pureHV}
ds^2=r^{\frac{-2\theta}{d}}\left(-r^{2z}dt^2+\frac{dr^2}{r^2}+r^2dx_i^2\right).
\end{equation}
Under the transformation
\begin{equation}
t\rightarrow\lambda^zt,\,\, x_i\rightarrow\lambda x_i,\,\,
r\rightarrow\lambda^{-1} r~, \label{scaling}
\end{equation}
with $z\neq 1$ indicating an anisotropy between
time and space. The metric (\ref{pureHV})
transforms as $ds\rightarrow
\lambda^{\theta/d}ds$ which breaks the
scale-invariance. A non-zero $\theta$, indicates
a hyperscaling violation in the dual field
theory. This metric is characterized by dynamical
critical exponent $z$ and  hyperscaling violation
exponent $\theta$ \cite{Fisher} and when
$\theta=0$ and $z\neq1$, it is reduced to the
Lifshitz metric
\cite{Kachru:2008,Taylor:2008,Pang:2009ad,Tarrio:2011},
while it describes the pure AdS metric when
$\theta=0$ and $z=1$.

Thermodynamically in these theories the entropy
scales as $T^{(d-\theta)/z}$, while in theories
with  gravity duals having the standard AdS
metric, the entropy scales as $T^{d}$. Note that
the hyperscaling violation leads to an effective
dimension $d_{\theta}=d-\theta$. It was addressed
in \cite{Ogawa:2011} that with the critical value
$d-\theta=1$, the entanglement entropy shows up a
logarithmic violation \cite{Wolf:2005}, and leads
to an infrared metric which holographically
represents  a compressible state with hidden
Fermi surfaces \cite{Huijse:2011ef}.

The introduction of Lifschitz scaling and
hyperscaling violation exponents in the metric of
the gravity sector has produced interesting
results. It was showed that these critical
exponents play important role in the retarded
Green's function in holographic systems with
finite charged density
\cite{Gursoy:2011,Alishahiha:201201,Alishahiha:201209}.
  It was found in \cite{Fang:2012} that for a specific
value of the critical exponent $z$, the
Luttinger's theorem is violated. In a dual
charged bulk theory with hyperscaling violation,
introducing a charge fermion as a probe to the
extremal gravity background, it was showed in
\cite{Alishahiha:201209} that the increase of the
Lifshitz factor $z$ and the hyperscaling factor
$\theta$ broadened and smoothed out the sharp
spectral function's peak, which  indicates that
the system does not have a Fermi surface.

In this work we will consider a dipole coupling in a charged
gravity bulk with hyperscaling violation and explore the spectral
function of the holographic dual fermion model. The dual model
with minimal coupling between fermionic field and gauge field was
discussed in our previous work \cite{Kuang:2014}, in  which we
showed that as the hyperscaling violation exponent is varied, the
fermionic system possesses Fermi, non-Fermi, marginal-Fermi and
log-oscillating liquid phases but failed to generate dynamically a
 gap.
In the case of Lifshitz geometry,  the generation
of Mott gap due to the dipole coupling  has been
observed in \cite{wu-lifshitz}.

 Our aim here is to study in details the behaviour of infrared (IR) and
ultra violet (UV) Green's functions in an attempt
to understand how the hyperscaling exponent
modifies the dipole effect on the formation of
Fermi surface, the liquid types of the low energy
excitations and the emergence of the Mott
insulating phase in these theories. Finally, we
will  show that there is a duality of the zeros
and the poles, first observed  in
\cite{Alsup:2014uca},
 in theories with hyperscaling violation.

The work  is organized as follows. In Section
\ref{SScalingGeometry}  we review the charged
black hole background with hyperscaling factor
and analyze the geometry in the near horizon
limit at zero temperature. We set up the
fermionic model and analytically study the
Green's function in the bulk theory in Section
\ref{SModel}. In Section \ref{SProperties} we
numerically investigate the properties of the UV
Green's function and we discuss the effect of the
hyperscaling exponent on the emergerence of the
gap, the formation of Fermi momentum and the
excitations due to the dipole coupling. Finally
in Section \ref{SConclusion} we present our
conclusions.

\section{The charged black branes with hyperscaling violation from Einstein-Maxwell-Dilaton theory}\label{SScalingGeometry}

In order to study the dipole coupling effects on the charged black
brane geometry with hyperscaling violation, we will consider the
$(3+1)$-dimensional Einstein-Maxwell-Dilaton action
\cite{Tarrio:2011}
\begin{equation}\label{EMDaction}
S_g=-\frac{1}{16\pi G}\int \mathrm{d}^{4}x \sqrt{-g}\left[R
-\frac{1}{2}(\partial \phi)^2+V(\phi)-
\frac{1}{4}\left(e^{\lambda_1\phi}F^{\mu\nu}F_{\mu\nu}+e^{\lambda_2\phi}\mathcal{F}^{\mu\nu}\mathcal{F}_{\mu\nu}\right)\right]~,
\end{equation}
which contains two $U(1)$ gauge fields coupled to
a  neutral scalar field $\phi$. The $U(1)$ field
$A$ with field strength $F_{\mu\nu}$ is necessary
to generate a charged black brane solution, while
the other gauge field $\mathcal{A}$ with field
strength $\mathcal{F}_{\mu\nu}$ is required to
generalize the geometry from AdS to the one with
the hyperscaling violation. Following
\cite{Alishahiha:201209} and our discussion in
\cite{Kuang:2014}, if one considers a potential
of the form
\begin{equation}\label{EMDpotential} V(\phi)=V_0e^{\gamma\phi}
\end{equation}
which is necessary  to obtain a general Lifshitz
form of the metric with hyperscaling violation,
one can find  the charged black brane solution
with hyperscaling violation
\cite{Alishahiha:201209}
\begin{eqnarray}\label{BGsolution}
&&
ds_{4}^2= r^{-\theta} \left(-r^{2z}f(r)dt^2+\frac{dr^2}{r^2f(r)}+r^2(dx^2+dy^2)\right),
\\
&&
\label{BGsolutionf}
f=1-\left(\frac{r_h}{r}\right)^{2+z-\theta}+\frac{Q^2}{r^{2(z-\theta+1)}}\left[1-\left(\frac{r_h}{r}\right)^{\theta-z}\right],
\\
&&
\mathcal{A}_t=-\slashed{\mu}r_h^{2+z-\theta}\left[1-\left(\frac{r}{r_h}\right)^{2+z-\theta}\right],
\\
&&
\label{RealAt}
A_t=\mu r_h^{\theta-z}\left[1-\left(\frac{r_h}{r}\right)^{z-\theta}\right],
\\
&&
\label{BGsolutionephi}
e^{\phi}=e^{\phi_0}r^{\sqrt{2(2-\theta)(z-1-\theta/2)}}.
\end{eqnarray}
where we have defined
\begin{eqnarray}\label{chemicalpontential}
&&
\slashed{\mu}=\frac{\sqrt{2(z-1)(2+z-\theta)}}{2+z-\theta}e^{\frac{2-\theta/2}{\sqrt{2(2-\theta)(z-1-\theta/2)}}\phi_0},
\\
&&
\label{Realchemicalpontential}
\mu=Q\sqrt{\frac{2(2-\theta)}{z-\theta}}e^{-\sqrt{\frac{z-1+\theta/2}{2(2-\theta)}}\phi_0}.
\end{eqnarray}

Here, $r_h$ is the radius of horizon satisfying $f(r_h)=0$ and
$Q=\frac{1}{16\pi G}\int e^{\lambda_1 \phi}F_{rt}$  is the total
charge of the black brane. All the parameters in the action,
dependent on the  Lifshitz scaling exponent $z$ and hyperscaling
violation exponents $\theta$, and they can be determined as
\begin{eqnarray}\label{parameters}
\lambda_1&=&\sqrt{\frac{2(z-1-\theta/2)}{2-\theta}}~,\nonumber\\
\lambda_2&=&-\frac{2(2-\theta/2)}{\sqrt{2(2-\theta)(z-\theta/2-1)}}~,\nonumber\\
\gamma&=&\frac{\theta}{\sqrt{2(2-\theta)(z-1-\theta/2)}}~,\nonumber\\
V_0&=&e^{\frac{-\theta\phi_0}{\sqrt{2(2-\theta)(z-1-\theta/2)}}}(z-\theta+1)(z-\theta+2)~.
\end{eqnarray}
The Hawking temperature of the black hole is
\begin{eqnarray}\label{temperature}
T=\frac{(2+z-\theta)r_h^z}{4\pi}\left[1-\frac{(z-\theta)Q^2}{2+z-\theta}r_h^{2(\theta-z-1)}\right].
\end{eqnarray}

Note that we have $z\geq 1$,  $\theta\geq 0$ and
the above solutions are not valid for $\theta=2$.
Before proceeding, we would like to remark more
on the parameters $z$ and $\theta$. First, the
background equations
~(\ref{BGsolution})-(\ref{BGsolutionephi}) are
valid only for $z\geq 1$ and $\theta\geq 0$. The
case of $z=1$ and $\theta=0$ corresponds to the
AdS geometry. Second, the condition $z-\theta\geq
0$ is required to make chemical potential
well-defined in the dual field theory. Third, it
is easy to see that $\theta< 2$ from
equation~(\ref{Realchemicalpontential}).
Combining the requirement of the null energy
condition
$(-\frac{\theta}{2}+1)(-\frac{\theta}{2}+z-1)\geq0$
\cite{Alishahiha:201209}, one can have
$\theta\leq2(z-1)$. Thus, in this charged
background, the range of the parameters is
\begin{eqnarray}\label{ParameterRegion}
\left\{
\begin{array}{rl}
&0\leq \theta \leq 2(z-1) \quad {\rm for} ~~~1\leq z<2   \ ,   \\
&0\leq\theta<2 \quad {\rm for}~~~ z\geq2  \ .
\end{array}\right.
\,
\end{eqnarray}

For convenience  we introduce the following rescaling
\begin{eqnarray}\label{rescaling}
&& r\rightarrow r_h r~,~~~t\rightarrow
\frac{t}{r_h^z}~,~~~(x,y)\rightarrow \frac{1}{r_h}(x,y)~,~~~
T\rightarrow
\frac{T}{r_h^z}~,
\nonumber\\
&& Q\rightarrow r_h^{(z-\theta+1)}Q~,~~~A_t\rightarrow r_h
A_t~,~~~\mathcal{A}_t\rightarrow r_h^{\theta-z-2} \mathcal{A}_t~,
\end{eqnarray}
and we set $r_h$ to be unity. Besides, we will
set $\phi_0=0$ in the following discussion
because it is an integration constant. Then we
can rewrite the metric factor $f(r)$ and the
gauge fields $\mathcal{A}_t$ and $A_t$  as
follows
\begin{eqnarray}\label{rescalingfr}
&&
f=1-\frac{1+Q^2}{r^{z+2-\theta}}+\frac{Q^2}{r^{2(z-\theta+1)}}~,
\\
&& \label{rescalingAtmathcal}
\mathcal{A}_t=-\slashed{\mu}\left[1-r^{2+z-\theta}\right]~,
\\
&& A_t=\mu \left[1-\left(\frac{1}{r}\right)^{z-\theta}\right]~,
\end{eqnarray}
and the dimensionless temperature in the form
\begin{eqnarray}\label{temperaturev1}
T=\frac{(2+z-\theta)}{4\pi}\left[1-\frac{(z-\theta)Q^2}{2+z-\theta}\right]~.
\end{eqnarray}
The zero-temperature limit can reached  when
$Q=\sqrt{\frac{2+z-\theta}{z-\theta}}$ and
$\mu=\frac{\sqrt{2(2-\theta)(2+z-\theta)}}{z-\theta}$.
Thus at zero temperature and in the $r\rightarrow r_h=1$ limit, we can reduce
\begin{eqnarray}\label{frTzerorh}
f(r)|_{T=0,r\rightarrow1}\simeq(z-\theta+1)(z-\theta+2)(r-1)^2\equiv\frac{1}{L_2^2}(r-1)^2.
\end{eqnarray}
Therefore, at the zero temperature, we obtain the
near horizon geometry $AdS_2\times \mathbb{R}^2$
with the  curvature radius $L_2\equiv
1/\sqrt{(z-\theta+1)(z-\theta+2)}$ of $AdS_2$
which depends explicitly on the  Lifshitz scaling
exponent $z$ and hyperscaling violation exponent
$\theta$. So, near the horizon, under the
transformation $r-1=\epsilon
\frac{L_{2}^{2}}{\varsigma}$ and
$t=\epsilon^{-1}\tau$  the metric and the gauge
fields are derived in the limit
$\epsilon\rightarrow 0$ with finite $\varsigma$
and $\tau$,
\begin{eqnarray} \label{MetricNearHorizon}
&&
ds^{2}=\frac{L_{2}^{2}}{\varsigma^{2}}(-d\tau^{2}+d\varsigma^{2})+dx^{2}+dy^2~,
\nonumber\\
&&
\mathcal{A}_{\tau}=\frac{\slashed{e}}{\varsigma}~,~~~~~A_{\tau}=\frac{e}{\varsigma}~,
\end{eqnarray}
where
$\slashed{e}=\slashed{\mu}(2+z-\theta)L_2^2$ and
$e=\mu(z-\theta)L_2^2$.

Before proceeding to the study of holographic fermionic,
we would like to give some comments on the geometry.
We rewrite the Einstein-Maxwell-Dilaton action (\ref{EMDaction}) as
\begin{equation}\label{EMDactionv1}
S_g=-\frac{1}{16\pi G}\int \mathrm{d}^{4}x \sqrt{-g}\left[R
-\frac{1}{4}Z(\phi)F^{\mu\nu}F_{\mu\nu}
-\frac{1}{2}(\partial \phi)^2+V_{eff}(\phi)
\right]~,
\end{equation}
where
\begin{eqnarray}\label{Veff}
Z(\phi)=e^{\lambda_1\phi}~~~and~~~V_{eff}(\phi)&=&V(\phi)
-\frac{1}{4}e^{\lambda_2\phi}\mathcal{F}^{\mu\nu}\mathcal{F}_{\mu\nu}=V_0e^{\gamma\phi}+V_{\mathcal{A}}e^{\gamma_{\mathcal{A}}\phi}
\end{eqnarray}
with
\begin{equation}
V_{\mathcal{A}}=\frac{1}{2}(z-1)(z+2-\theta)~~~and~~~
\gamma_{\mathcal{A}}=\frac{\theta}{\sqrt{2(2-\theta)(z-1-\theta/2)}}
~.
\end{equation}

It is easy to check that $\theta=0$ and $z=1$, leading to
$Z(\phi)=1$ and $V_{eff}(\phi)=6$, give us the RN-AdS black hole
solution. When
$V_{eff}(\phi)=V_0+V_{\mathcal{A}}e^{\gamma_{\mathcal{A}}\phi}$,
it was addressed in \cite{Tarrio:2011} that the action admits a
Lifshitz black hole solution with $\theta=0$. Therefore, the
effective potential $V_{eff}(\phi)$ controls the UV geometry,
which can be changed from AdS to Lifshitz, even to the geometry
with hyperscaling violation accompanying the parameters
$\lambda_1$ and $\gamma_{\mathcal{A}}$.

The effective potential (\ref{Veff}) is very different from that
shown in \cite{Gouteraux:2011,Charmousis:2010}, in which the
authors constructed an effective holographic theory with a scalar
field $\phi$ and only one gauge field $A_\mu$. In their work, the
parameters $\lambda_1$ in the gauge coupling function
$Z(\phi)=e^{\lambda_1\phi}$ and $\gamma$ in the potential
$V(\phi)=V_0e^{\gamma\phi}$ control the IR behavior. While for the
UV limit, the potential $V(\phi)$ is required to vanish so that it
usually leads to asymptotically AdS. When the gauge field in
\cite{Gouteraux:2011,Charmousis:2010} plays the role of
$\mathcal{F}_{\mu\nu}$ but not that of $F_{\mu\nu}$ in the action
(\ref{EMDaction}), one can obtain UV geometry with asymptotically
Lifshitz-AdS but uncharged as discussed in \cite{Taylor:2008}.

\section{The holographic fermionic model}\label{SModel}

\subsection{The Dirac equation}

To probe the geometry with hyperscaling violation, we consider the
following Dirac action including a dipole moment coupling  between
the fermion and the gauge field\footnote{The Dirac action (\ref{actionspinor})
depends on the effective chemical potential $\mu_{eff}\equiv \mu q$ and the product of the dipole coupling $p$ and $\mu$.
That is to say, the Dirac action depends on the combination of $g_F q$ with $g_F=\frac{2}{\sqrt{Z(\phi)}}$.
For the case of $z=1$ and $\theta=0$, $g_F q=2$ because of $Z(\phi)=1$.
It is different from Refs.\cite{Liu:2009,Edalati:2010a,Edalati:2010b}, in which the authors set $g_F=1$.
Therefore, we remind readers to note that the charge $q$ and the bulk dipole coupling $p$ for $z=1$ and $\theta=0$ here
will correspond to $q/2$ and $p/2$ in Refs.\cite{Liu:2009,Edalati:2010a,Edalati:2010b}.}
\cite{Edalati:2010a,Edalati:2010b}
\begin{eqnarray}
\label{actionspinor} S_{D}=i\int d^{d+1}x
\sqrt{-g}\,\overline{\zeta}\left(\Gamma^{a}\mathcal{D}_{a} - m -i
p \slashed{F}\right)\zeta,
\end{eqnarray}
where
$\mathcal{D}_{a}=\partial_{a}+\frac{1}{4}(\omega_{\mu\nu})_{a}\Gamma^{\mu\nu}-iqA_{a}$
and
$\slashed{F}=\frac{1}{2}\Gamma^{\mu\nu}(e_\mu)^a(e_\nu)^bF_{ab}$
with $\Gamma^{\mu\nu}=\frac{1}{2}[\Gamma^\mu,\Gamma^\nu]$ and the
spin connection
$(\omega_{\mu\nu})_{a}=(e_\mu)^b\nabla_a(e_\nu)_b$. Here, it is
worthwhile to point out that as a ``bottom-up" approach, one can
also add an additional dipole coupling term between the Dirac
field and the other gauge field $\mathcal{A}$, i.e.,
$\tilde{p}\overline{\zeta}\slashed{\mathcal{F}}\zeta$ with
$\slashed{\mathcal{F}}=\frac{1}{2}\Gamma^{\mu\nu}(e_\mu)^a(e_\nu)^b\mathcal{F}_{ab}$,
into the Dirac action (\ref{actionspinor}) and then explore its
effects on the spectral  function as well as the
enhance/competition between $p$ and $\tilde{p}$.
However, $\tilde{p}\overline{\zeta}\slashed{\mathcal{F}}\zeta$ will be divergent in the UV boundary.
Therefore, a new boundary counterterm is usually needed to obtain a finite on-shell action.
Ref.\cite{0912.1061} can shed a light on how to understand the divergences
from the point of view of the field theory and the boundary counterterms.
We shall further explore this subject in the future.
In this work, we will only investigate the case with $\tilde{p}=0$.

Now, we shall derive the Dirac equation.
From the above action, with the redefinition
$\zeta=(-g g^{rr})^{-\frac{1}{4}}\mathcal{F}$ and
a Fourier transformation $\mathcal{F}=F
e^{-i\omega t +ik_{i}x^{i}}$, we can write the
Dirac equation in the Fourier space
\begin{eqnarray}
\label{DiracEinFourier} (\sqrt{g^{rr}}\Gamma^{r}\partial_{r}- m)F
-i(\omega+q A_{t})\sqrt{g^{tt}}\Gamma^{t}F +i( k \sqrt{g^{xx}}-p
\sqrt{g^{tt}g^{rr}}\partial_{r}A_{t})\Gamma^{x}F =0~.
\end{eqnarray}
Due to the rotational symmetry in $x-y$ plane, we have set $k_{x}=k$ and $k_{y}=0$.
With the choice of the usual gamma matrices
\begin{eqnarray}
\label{GammaMatrices}
 && \Gamma^{r} = \left( \begin{array}{cc}
-\sigma^3 & 0  \\
0 & -\sigma^3
\end{array} \right), \;\;
 \Gamma^{t} = \left( \begin{array}{cc}
 i \sigma^1 & 0  \\
0 & i \sigma^1
\end{array} \right),  \;\;
\Gamma^{x} = \left( \begin{array}{cc}
-\sigma^2 & 0  \\
0 & \sigma^2
\end{array} \right),
\qquad \ldots
\end{eqnarray}
The Dirac equation takes the form
\begin{eqnarray} \label{DiracEF}
\left[(\partial_{r}+m\sqrt{g_{rr}}\sigma^3)
-\sqrt{\frac{g_{rr}}{g_{tt}}}(\omega+qA_{t})i\sigma^2 -((-1)^{I} k
\sqrt{\frac{g_{rr}}{g_{xx}}}-p
\sqrt{g^{tt}}\partial_{r}A_{t})\sigma^1 \right] F_{I} =0~
\end{eqnarray}
with $I=1,2$. Decomposing $F_{I}$ into $F_{I}=(\mathcal{A}_{I},\mathcal{B}_{I})^{T}$, we can get the equation of motion for all the component as
\begin{eqnarray} \label{DiracEAB1}
&& (\partial_{r}+m\sqrt{g_{rr}})\mathcal{A}_{I}
-\sqrt{\frac{g_{rr}}{g_{tt}}}(\omega+qA_{t})\mathcal{B}_{I}
-((-1)^{I} k \sqrt{\frac{g_{rr}}{g_{xx}}}-p
\sqrt{g^{tt}}\partial_{r}A_{t})\mathcal{B}_{I} =0~,
\\
&& \label{DiracEAB2} (\partial_{r}-m\sqrt{g_{rr}})\mathcal{B}_{I}
+\sqrt{\frac{g_{rr}}{g_{tt}}}(\omega+qA_{t})\mathcal{A}_{I}
-((-1)^{I} k \sqrt{\frac{g_{rr}}{g_{xx}}}-p
\sqrt{g^{tt}}\partial_{r}A_{t}) \mathcal{A}_{I} =0~.
\end{eqnarray}
Then defining $\xi_{I}\equiv
\frac{\mathcal{A}_{I}}{\mathcal{B}_{I}}$ , we can
obtain the flow equation
\begin{eqnarray} \label{DiracEF1}
(\partial_{r}+2m\sqrt{g_{rr}}) \xi_{I}
-\left[ v_{-} + (-1)^{I} k \sqrt{\frac{g_{rr}}{g_{xx}}}  \right]
- \left[ v_{+} - (-1)^{I} k \sqrt{\frac{g_{rr}}{g_{xx}}}  \right]\xi_{I}^{2}
=0
~,
\end{eqnarray}
where we have defined
$v_{\pm}=\sqrt{\frac{g_{rr}}{g_{tt}}}(\omega+q
A_{t})\pm p \sqrt{g^{tt}}\partial_{r}A_{t}$. For
the convenience of numerical calculation later,
we  make a transformation $r=1/u$, so that the
flow equation (\ref{DiracEF1}) can be rewritten
as
\begin{eqnarray} \label{DiracEF2}
\left(\sqrt{f}\partial_{u}-2m u^{\frac{\theta}{2}-1}\right) \xi_{I}
+\left[ \frac{\tilde{v_{-}}}{u} + (-1)^{I} k  \right]
+ \left[ \frac{\tilde{v_{+}}}{u} - (-1)^{I} k  \right]\xi_{I}^{2}
=0
~
\end{eqnarray}
with
\begin{eqnarray}\label{titdev}
\tilde{v_{\pm}}=\frac{u^z}{\sqrt{f}}(\omega+qA_t)\mp p u^{z-\theta/2+1}\partial_{u}A_{t}.
\end{eqnarray}

\subsection{Green's function}
\subsubsection{UV limit}
We first consider the UV limit of the Dirac
equation.
 Equation (\ref{pureHV}) with hyperscaling violation gives  $g_{rr}=r^{-\theta-2}$,
$g_{tt}=r^{2z-\theta}$ and $g_{xx}=g_{yy}=r^{2-\theta}$ in the UV
limit. Therefore, the Dirac equation (\ref{DiracEF}) becomes
\begin{eqnarray}\label{DiracEFHV}
\left[\partial_r+mr^{-\frac{\theta}{2}-1}\sigma^3-r^{-1-z}(\omega+q\mu)i\sigma^2-((-1)^Ikr^{-2}
-p\mu(z-\theta)r^{\frac{3}{2}\theta-2z-1})\sigma^1 \right]F_I =0~.
\end{eqnarray}
Considering the allowed range of values of $\theta$ and $z$ in
equation (\ref{ParameterRegion}),  we can reduce equation
(\ref{DiracEFHV}) in the limit of $r\rightarrow \infty$ to
\begin{eqnarray} \label{thetaB2}
\left(\partial_{r}+\frac{m}{r^{\frac{\theta}{2}+1}}\sigma^3\right)F_{I}
\approx 0~.
\end{eqnarray}
For $\theta=0$, equation (\ref{thetaB2}) give the following solution
\begin{eqnarray}\label{FIbIaIzerom}
F_{I} \buildrel{r \to \infty}\over {\approx} b_{I}r^{m}\left( \begin{matrix} 0 \cr  1 \end{matrix}\right)
+a_{I}r^{-m}\left( \begin{matrix} 1 \cr  0 \end{matrix}\right),
\end{eqnarray}
which agrees well with the results in AdS or Lifshitz-AdS geometry.
While for $\theta\neq0$, the asymptotical behavior of $F_{I}$ becomes subtle.
Firstly, equation (\ref{thetaB2}) for $\theta\neq 0$ satisfies the solutions
\begin{eqnarray}
\mathcal{A}_I=a_Ie^{\frac{2m}{\theta}r^{-\frac{\theta}{2}}}\simeq a_I\left(1+\frac{2m}{\theta}r^{-\frac{\theta}{2}}+\ldots\right),~~~~
\mathcal{B}_I=b_Ie^{-\frac{2m}{\theta}r^{-\frac{\theta}{2}}}\simeq b_I\left(1-\frac{2m}{\theta}r^{-\frac{\theta}{2}}+\ldots\right).
\end{eqnarray}
Due to $\frac{2m}{\theta}r^{-\frac{\theta}{2}}\rightarrow 0$ for $\theta> 0$ in the limit of $r\rightarrow\infty$,
we can deduce that at the leading order, the behavior of $F_I$ is
\begin{eqnarray}\label{FIbIaI}
F_{I} \buildrel{r \to \infty}\over {\approx} b_{I}\left( \begin{matrix} 0 \cr  1 \end{matrix}\right)
+a_{I}\left( \begin{matrix} 1 \cr  0 \end{matrix}\right),
\end{eqnarray}
which is the same asymptotical behavior as the case of zero mass in AdS (Lifshitz-AdS) geometry.
According to the discussion in \cite{Faulkner:2009}, it applies that we can choose either $a_{I}$ or $b_{I}$
as the source when we quantize Fermi field with different boundary conditions. In this work, we will choose
$b_{I}$ as the source and $a_{I}$ as the response.
Thus, in the regime of linear response, the boundary Green's functions can be extracted by $G_{I}=\frac{a_I}{b_I}$.
If we define
\begin{eqnarray}
\xi_{I}\equiv
\frac{\mathcal{A}_{I}}{\mathcal{B}_{I}}.~
\end{eqnarray}
So the boundary Green's functions can be expressed in terms of
$\xi_I$
\begin{eqnarray} \label{GreenFBoundary}
G (\omega,k)=\left( \begin{array}{cc}
G_{1}   & 0  \\
0  & G_{2} \end{array} \right)=
\left\{
\begin{array}{rl}
&\lim_{r\rightarrow \infty}
r^{2m}
\left( \begin{array}{cc}
\xi_{1}   & 0  \\
0  & \xi_{2} \end{array} \right),
~~~~~~\theta=0   \ ,   \\
&\lim_{r\rightarrow \infty}
\left( \begin{array}{cc}
\xi_{1}   & 0  \\
0  & \xi_{2} \end{array} \right),~~~~~~
\theta>0 \ .
\end{array}\right.
\end{eqnarray}

Also, from equation (\ref{DiracEF2}), we can see that the Green
function has  the following symmetry
\begin{equation}\label{Gsym}
G_{1}(\omega,k;m,p)=G_{2}(\omega,-k;m,p)~.
\end{equation}

\subsubsection{IR limit}

We then turn to the IR limit of the equations of motion. Since the
near horizon geometry is $AdS_2\times \mathbb{R}^2$, we will use
the metric (\ref{MetricNearHorizon}). In the low energy frequency
limit $\omega\rightarrow 0$, the Dirac equation takes the form
\begin{eqnarray} \label{DiracEAdS2}
\varsigma \partial_{\varsigma}F_{I}-\left[m
L_{2}\sigma^{3}+((z-\theta)p\mu-(-1)^{I}k)L_{2}\sigma^{1}
-i\sigma^{2}q(z-\theta)\mu L_2^2\right]F_{I}=0~.
\end{eqnarray}
Note that  we also choose the same Gamma matrices
(\ref{GammaMatrices})  but change
$\Gamma^{\varsigma}=-\Gamma^{r}$ to reflect the
orientation between the coordinates $r$ and
$\varsigma$. As it was discussed in
\cite{Faulkner:2009}, equation (\ref{DiracEAdS2})
coincides with the equation of motion for spinor
fields in $AdS_2$ background with masses
\begin{equation}\label{mtilde}
[m,\tilde{m}_{I}=(z-\theta)p\mu-(-1)^{I}k]~,
\end{equation}
where $\tilde{m}_{I}(I=1,2)$ are time-reversal violating mass
terms. Then $F_{I}^{(0)}(\varsigma)$ is dual  to the spinor
operators $\mathbb{O}_{I}$ in the IR $CFT_{1}$ and  the conformal
dimensions of the operators are
$\delta_{I}=\nu_{I}(k)+\frac{1}{2}$ with
\begin{equation}\label{nuk}
\nu_{I}(k)=\sqrt{(m^2+\tilde{m}_{I}^2)L_{2}^2-\big[(z-\theta)q\mu
L_{2}^2\big]^2}~~~~~~~(I=1,2)~.
\end{equation}
In (\ref{nuk})  the Lifshitz and hyperscaling
violation exponents as well as the coupling
parameters like the dipole moment appear
explicitly and they imprint their scalings in the
IR limit.

\begin{figure}
\center{
\includegraphics[scale=0.45]{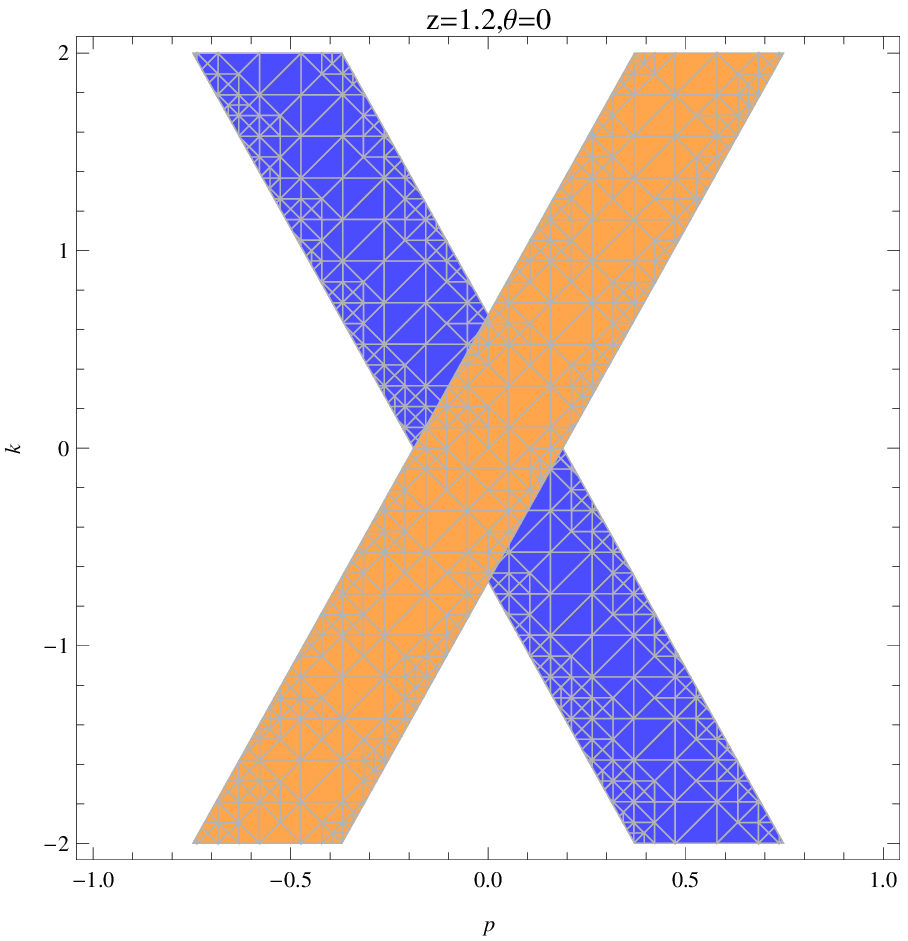}\hspace{0.5cm}
\includegraphics[scale=0.45]{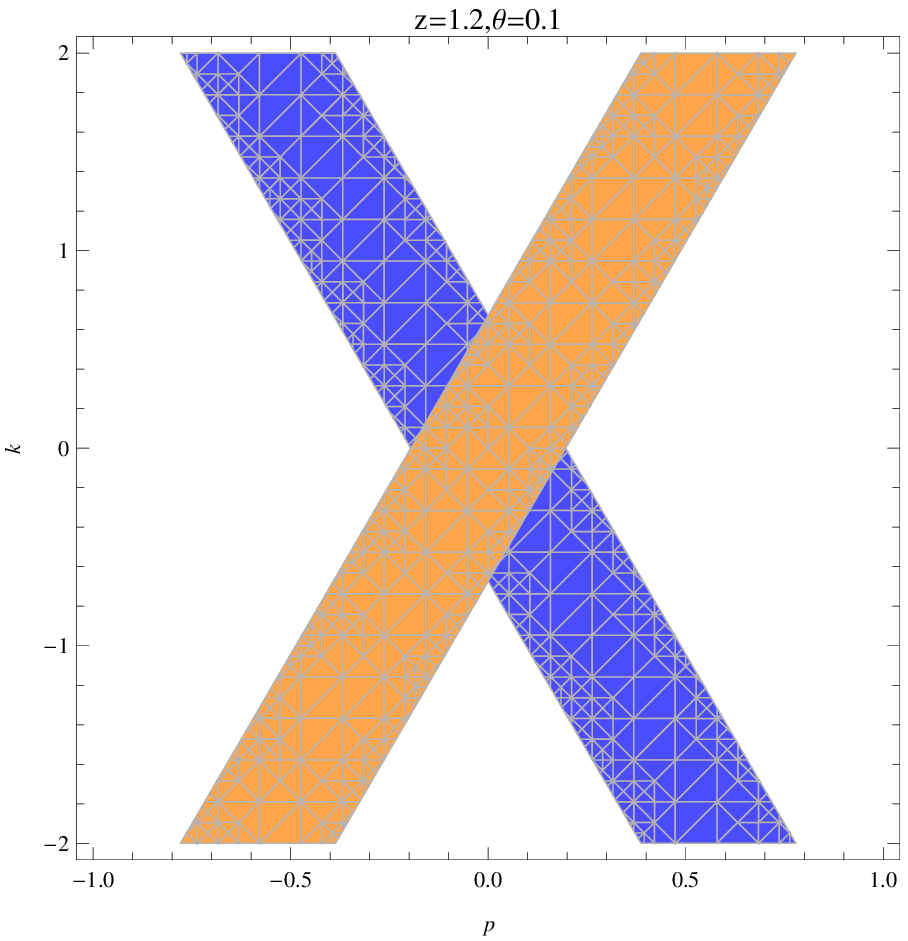}\hspace{0.5cm}
\includegraphics[scale=0.45]{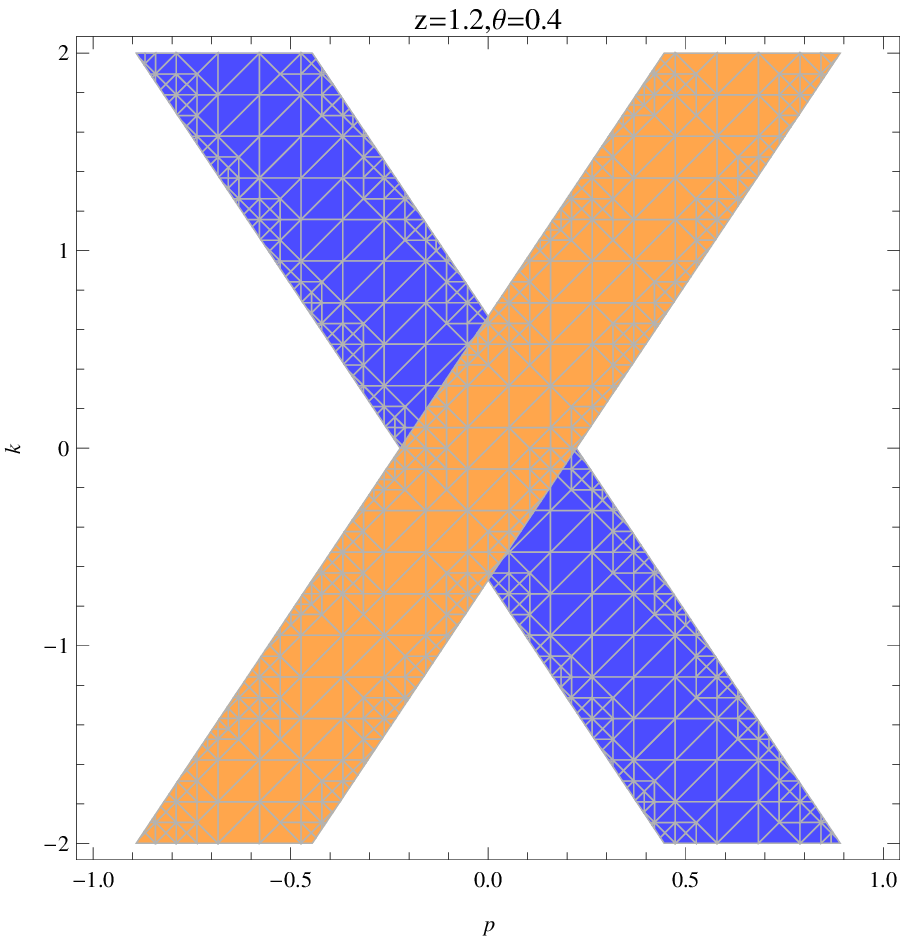}\hspace{0.5cm}
\caption{\label{figoscillation} The oscillatory regions for $z=1.2;\theta=0$(left),
$z=1.2;\theta=0.1$(middle) and $z=1.2;\theta=0.4$(right). The orange region denotes the
oscillatory region $\mathfrak{I}_{2}$ for $G_{22}(\omega, k)$ while the blue region corresponds
to oscillatory region $\mathfrak{I}_{1}$ for $G_{11}(\omega, k)$.}}
\end{figure}

 There exists a range of momentum
\begin{equation}\label{kosc}
k\in \mathfrak{I}_{I}=\left[(-1)^{I}(z-\theta)p\mu-(z-\theta)q\mu L_2,(-1)^{I}(z-\theta)p\mu+(z-\theta)q\mu L_2\right]
\end{equation}
in which $\nu(k)$ becomes pure imaginary. This
region of momentum space is considered as the
oscillatory region. If the Fermi momentum falls
in this region, the peak will lose its meaning as
a Fermi surface \cite{Faulkner:2009}. From the
expression (\ref{kosc}), it is obvious that for
fixed $z$ and $\theta$, the oscillatory regions
for the two-dimensional dual operator are
coincident with
$\mathfrak{I}_{1}=\mathfrak{I}_{2}=\left[-(z-\theta)q\mu
L_2,(z-\theta)q\mu L_2\right]$ for minimal
coupling $p=0$, but they will separate when we
turn on the dipole coupling.  And when the dipole
coupling satisfies $\mid p \mid >qL_2$, they will
have no intersection.   The separation behavior
of the regimes $\mathfrak{I}_{1}$  and
$\mathfrak{I}_{2}$ versus $p$ for various
exponents can be seen in
Fig.~\ref{figoscillation}. It is obvious that the
symmetry Eq.~(\ref{Gsym}) is well embodied in the
figure, i.e.,
$\mathfrak{I}_{1}=-\mathfrak{I}_{2}$ for fixed
dipole coupling. Another property we can see from
Eq.~(\ref{kosc}) and the Fig.1 is that
$\mathfrak{I}_{1}$ at $p$ coincides with
$\mathfrak{I}_{2}$ at $-p$. The figure also shows
that for larger hyperscaling exponent, the
boundary of the oscillatory region is more
smooth. We will see later that this behaviour
will reflect  a phase transition which occurs at
a critical dipole coupling  with the variation of
the exponents.

Also following the matching method proposed in \cite{Faulkner:2009}, we can match the inner and outer solutions in the matching region
($\varsigma\rightarrow0$ and $\omega/\varsigma\rightarrow0$). We express the coefficients $a_{I}$ and $b_{I}$ in (\ref{FIbIaI}) as
\begin{eqnarray}\label{coefficients}
a_{I}&=&[a_{I}^{(0)}+\omega a_{I}^{(1)}+\cdots]+[\tilde{a}_{I}^{(0)}+\omega \tilde{a}_{I}^{(1)}+\cdots] \mathcal{G}_{I}(k,\omega)~,\nonumber \\
b_{I}&=&[b_{I}^{(0)}+\omega
b_{I}^{(1)}+\cdots]+[\tilde{b}_{I}^{(0)}+\omega
\tilde{b}_{I}^{(1)}+\cdots] \mathcal{G}_{I}(k,\omega)~,
\end{eqnarray}
where coefficients $a_{I}^{(n)},
\tilde{a}_{I}^{(n)}, b_{I}^{(n)}$ and
$\tilde{b}_{I}^{(n)}$ are to be determined and
\begin{equation}\label{greenads2}
 \mathcal{G}_{I}(k,\omega)=\left\{e^{-i \pi \nu_{I}(k)}\frac{\Gamma(-2\nu_{I}(k))\Gamma(1+\nu_{I}(k)-i(z-\theta)q\mu L_2^2)
 [(m+i\tilde{m}_{I})L_2-i(z-\theta)q\mu L_2^2-\nu_{I}(k)]}
{\Gamma(2\nu_{I}(k))\Gamma(1-\nu_{I}(k)-i(z-\theta)q\mu L_2^2)[(m+i\tilde{m}_{I})L_2-i(z-\theta)q\mu L_2^2
+\nu_{I}(k)]}\right\}\omega^{2  \nu_{I}(k)}
\end{equation}
is the retarded Green functions of the dual
operators $\mathbb{O}_{I}$. We see that the
hyperscaling violation exponent and dipole
coupling explicitly modify the boundary  Green's
function. More discussions on the  above Green's
function can be seen in \cite{Faulkner:2009}
where it was found  that (\ref{greenads2}) is
only valid when $2\nu_{I}(k)$ is not an integer.
In the case when it is an integer, terms like
$\omega^n log(\omega)$ should be added.

Since the IR geometry of the charged geometry with hyperscaling
violation  is also $AdS_2\times \mathbb{R}^2$ as that in RN-AdS
black brane and Lifshitz AdS black brane, we can easily derive the
boundary conditions of $\xi$ at the horizon $r_h=1$ for
$\omega\neq 0$ and $\omega=0$ as
\begin{eqnarray}\label{HorizonCondition}
\left\{
\begin{array}{rl}
&\xi_{I}\buildrel{r \to 1}\over=i ~~~for~~~\omega\neq 0   \ ,   \\
&\xi_{I}\buildrel{r \to 1}\over =\frac{mL_2-\nu_{I}(k)}
{(z-\theta)q\mu L_2^2+\tilde{m}_{I}L_2}~~~for~~~\omega=0 \ .
\end{array}\right.
\,
\end{eqnarray}

\section{The effect of dipole coupling  on the spectral function in the Fermionic system }\label{SProperties}
We numerically solve the flow equation
(\ref{DiracEF2}) and read off the asymptotic
values to  extract the retarded Green functions.
By studying the spectral function, we will
explore the generation of a gap phase due to
large enough dipole coupling in the bulk with
different hyperscaling violation strength. Then,
we also go to small  dipole coupling to find out
the Fermi momentum and the type of excitations
near the Fermi surface. Our study will focus on
taking $m=0$ and $q=0.5$.

\subsection{The emergence of the Mott gap}

We will calculate the density of states
$A(\omega)$ by doing the integration of the Fermi
spectral function $A(\omega; k)=\rm{Tr}
[\rm{Im}G(\omega;k)]$ over $k$. The  effects of
the dynamical Lifshitz exponent on the
holographic fermionic systems and the emergence
of the Mott gap were discussed in
\cite{wu-lifshitz}. It was found that the gap
opens easier for the larger Lifshitz exponent.
Here we will mainly focus on the effects of the
hyperscaling exponent.

\begin{figure}
\center{
\includegraphics[scale=0.21]{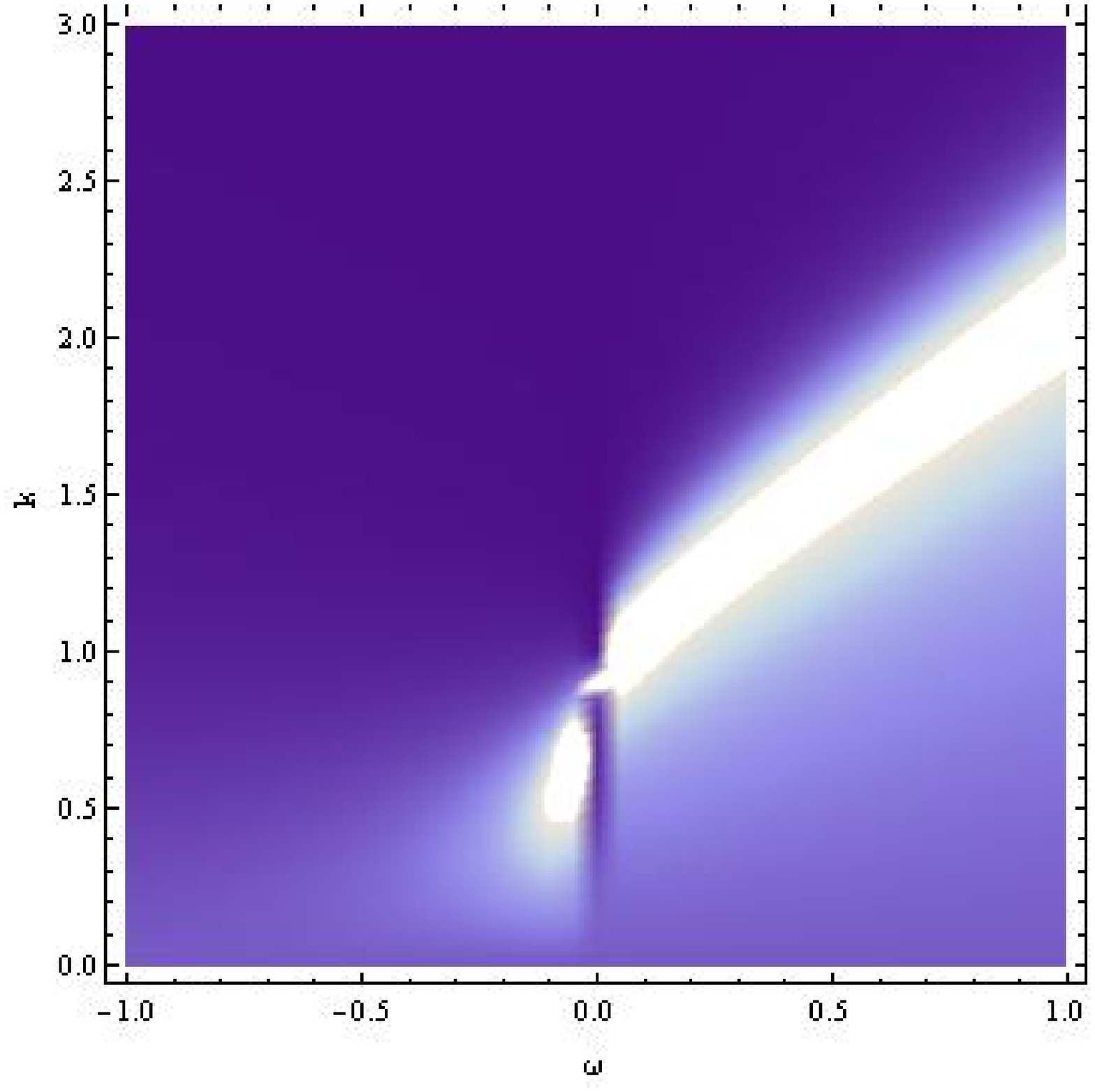}\hspace{1.5cm}
\includegraphics[scale=0.21]{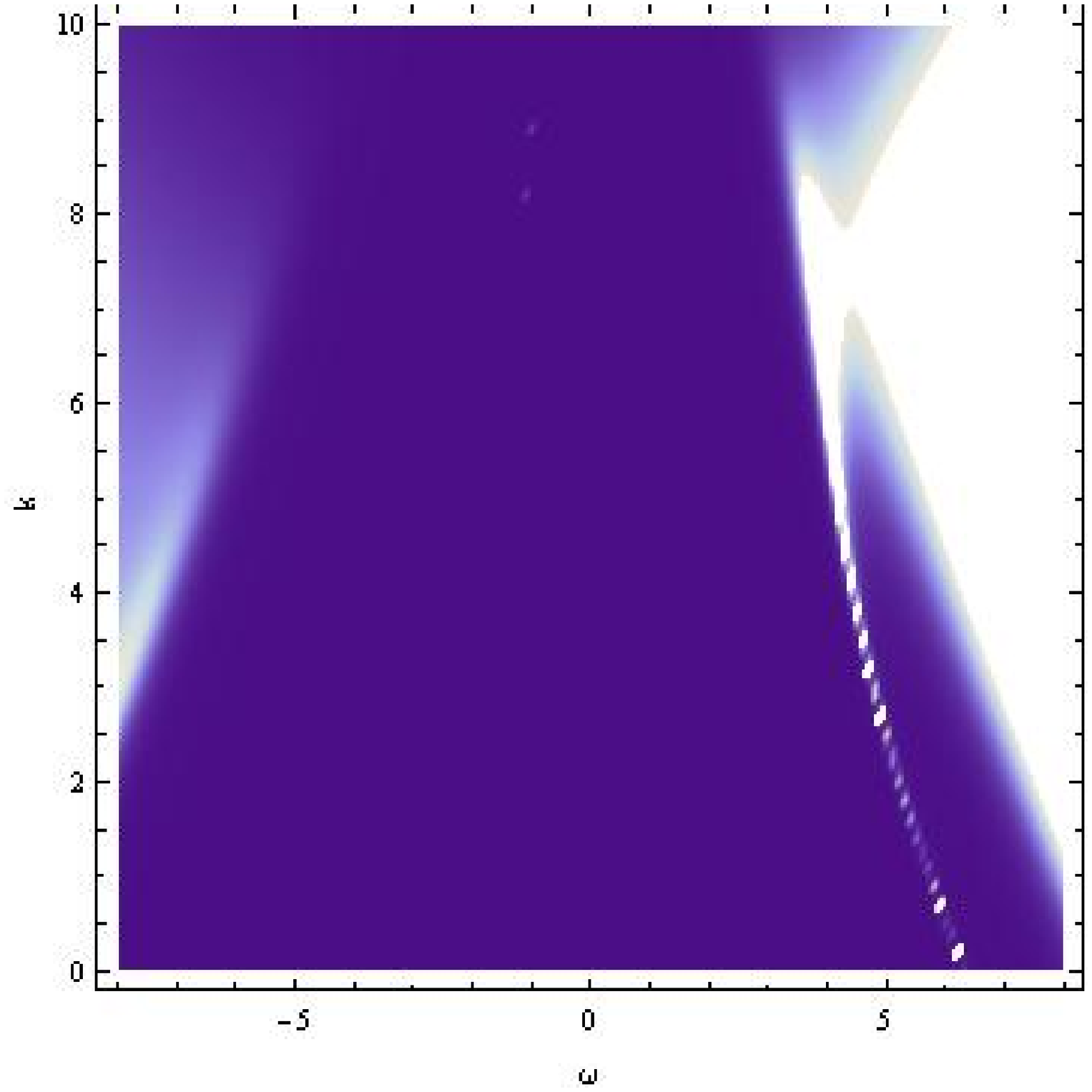}\\
\caption{\label{fig3Dz1c0} $\rm{Im} [G_{22}(\omega,k)]$ for $p=0$ (left plane) and $p=6$ (right plane) in RN-AdS background.}}
\end{figure}

For comparison, we first show the results of the
gap emergence  due to the dipole coupling in a
RN-AdS black hole background, i.e., with $z=1$
and $\theta=0$. In Fig.~\ref{fig3Dz1c0}, we show
the results of $\rm{Im} G_{22}$ for $p=0$ and
$p=6$. The left plot is for $p=0$, where the
quasi-particle-like peak at $\omega=0$ indicates
a Fermi surface near which the low energy
excitation is non-Fermi liquid type
\cite{Liu:2009}. The right plot is for taking
$p=6$ where an explicit gap is around $\omega=0$.
This is the dipole coupling effect first studied
in \cite{Edalati:2010a,Edalati:2010b}.
\begin{figure}
\center{
\includegraphics[scale=0.21]{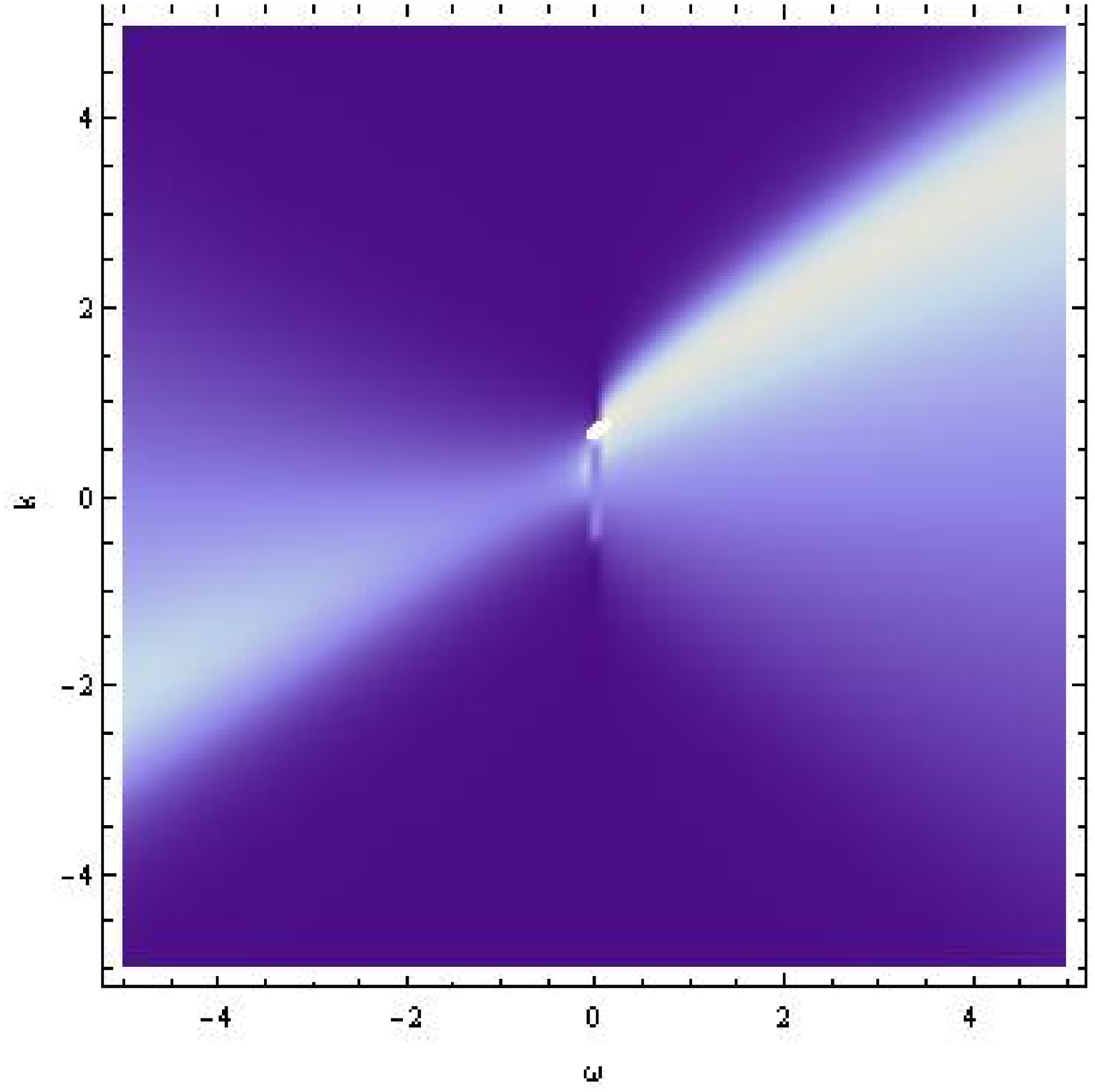}\hspace{1.5cm}
\includegraphics[scale=0.21]{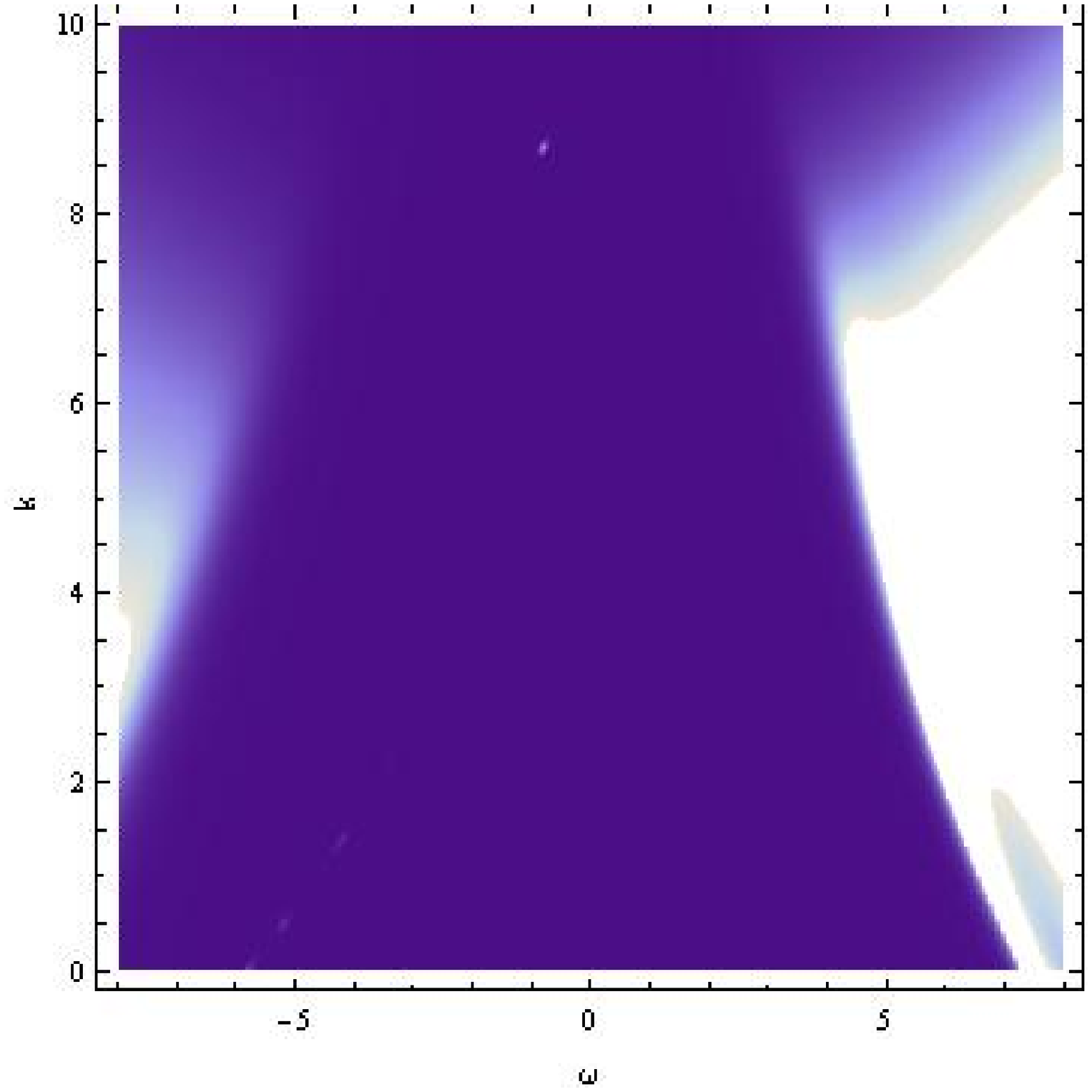}\\
\includegraphics[scale=0.21]{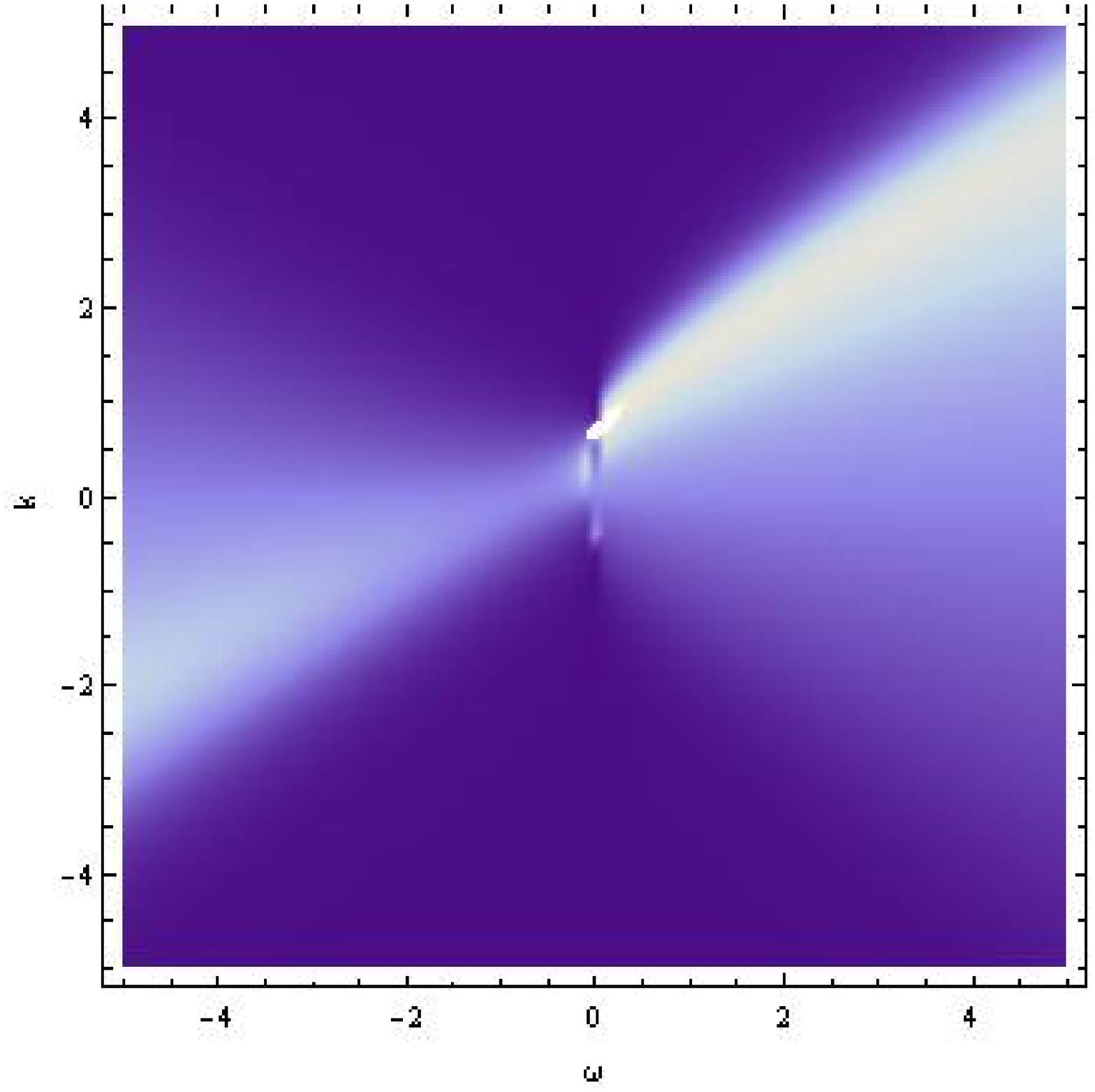}\hspace{1.5cm}
\includegraphics[scale=0.21]{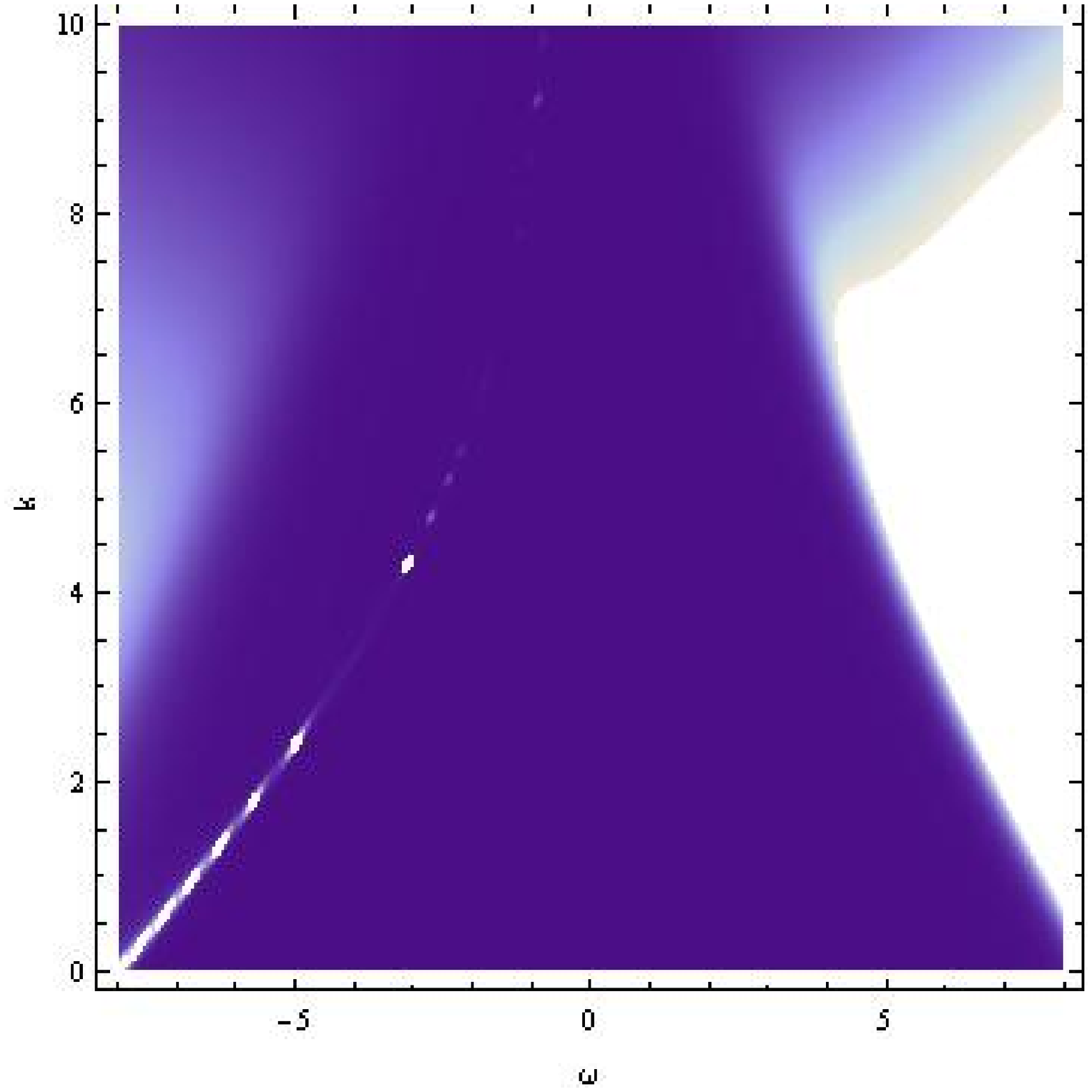}\\
\caption{\label{fig3Dz12c01c04} The density plots of $\rm{Im}
[G_{22}(\omega,k)]$ for $p=0$  (left plane) and $p=6$ (right
plane). The exponents for the up plane are  $z=1.2$ and
$\theta=0.1$ while the bottom plane are for  $z=1.2$ and
$\theta=0.4$.}}
\end{figure}
\begin{figure}
\center{
\includegraphics[scale=0.7]{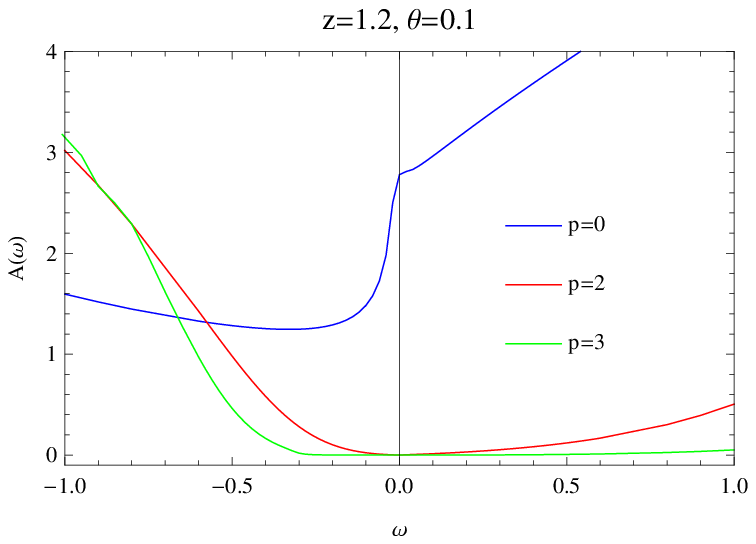}\hspace{0.5cm}
\includegraphics[scale=0.7]{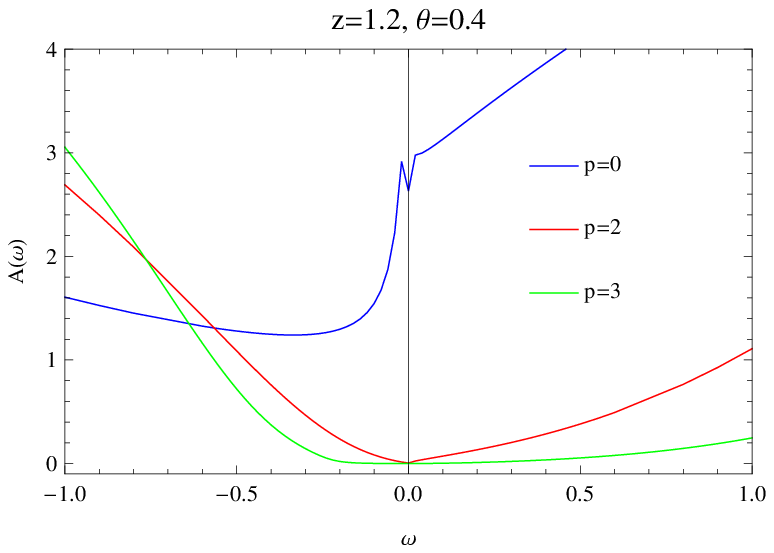}
\caption{\label{figAw} The behaviour of spectral function
$A(\omega)$ with $\omega$ for $z=1.2$.  We set $\theta=0.1$ (left
plane) and $\theta=0.4$ (right plane) respectively.}}
\end{figure}
\begin{figure}
\center{
\includegraphics[scale=0.7]{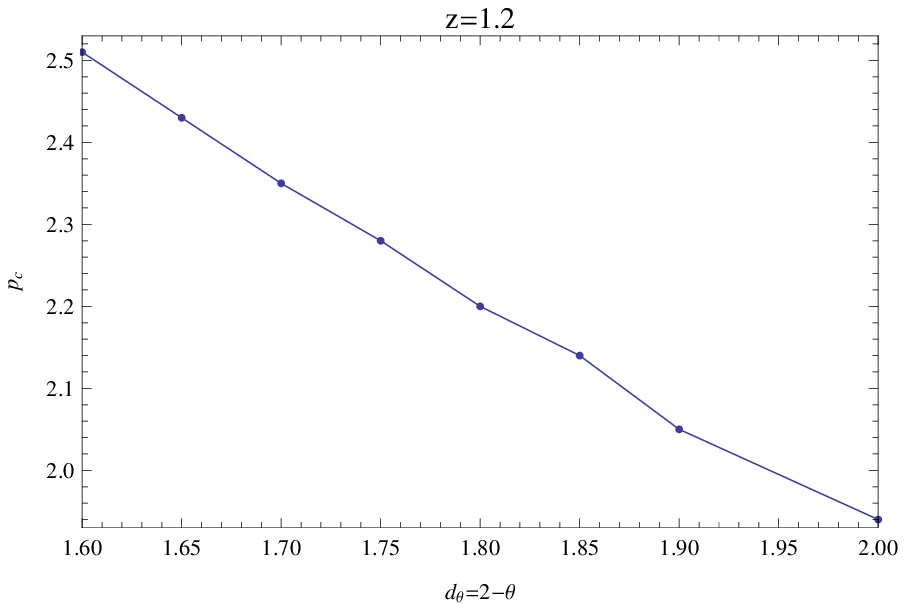}\hspace{0.5cm}
\caption{\label{figpc} The dependence of $p_c$ for the gap opening
on the effective  dimension $d_{\theta}=d-\theta$ with $z=1.2$.
Here in our case we have $d=2$ .}}
\end{figure}

Then we turn on hyperscaling violation $\theta$
to consider its effect on the dual  field theory.
The Green functions with fixed $z=1.2$ for
different $\theta$ are shown in
Fig.~\ref{fig3Dz12c01c04}. Comparing the plots in
Fig.~\ref{fig3Dz12c01c04}, we find that for a
fixed value of the dipole coupling, the larger
hyperscaling violation exponent introduces
smaller gap, which implies that in the theory
with hyperscaling violation the Mott gap phase is
hard to be formed. This feature is explicitly
shown in Fig.~\ref{figAw}, where we present the
density of states near the chemical potential
with $p$ by changing the hyperscaling violation
exponent. Each plot shows that as the dipole
coupling is strengthened, the spectral function
will be suppressed near the zero frequency, then
a gap will open at some critical value,
accompanying spectral weight transfer from
positive frequency band to negative band.

It is important to note that in
Fig.~\ref{fig3Dz12c01c04} an explicit band is
shown to be generated  which  disperses as $k$
increases.   This band is very important because
it  contributes to the spectral function,
especially the spectral weight transfer at
negative enough frequency. The mechanism of the
generation of this band
 is still called for.

Carrying out some exact  calculations to define
the gap when the spectral function is below $\sim
0.0001$, we determine that the gap opens at
$p_c\simeq 1.94$ for $z=1.2$ and $\theta=0$,
$p_c\simeq 2.05$ for $z=1.2$ and $\theta=0.1$,
and $p_c\simeq 2.51$ for $z=1.2$ and
$\theta=0.4$. These results show that as $\theta$
increases, $p_c$ becomes larger, which means that
the gap is more difficult to be generated
dynamically.

This is a very interesting result. While the
anisotropic Lifshitz scaling tends to decrease
the critical value of the dipole moment $p_c$,
the increase of Lifshitz exponent can make the
Mott gap open up more easily\cite{wu-lifshitz}.
However, the increase of the hyperscaling
violation factor plays  the opposite role, which
 makes it harder for the Mott insulating phase to
appear. This behaviour can be understood as
follows. As we discussed in the introduction, the
hyperscaling violation factor introduces an
effective dimension $d_\theta=d-\theta$ into the
theory with $d=2$ in our holographic model. So
larger hyperscaling violation factor corresponds
to lower effective dimension. Then if one looks
at the flow equation, the  critical $p$ is larger
for lower dimension, because the spacetime
dimension compensates the effect of $p$. This
behaviour was observed for the first time in
\cite{Kuang:2012}. Furthermore, we get the
dependence of $p_c$ for the gap opening on the
effective dimension $d_{\theta}$ with $z=1.2$ in
Fig.~\ref{figpc}. It seems that in systems with
lower effective dimension  the Mott gap phase is
harder to emerge.

\subsection{The formation of the Fermi surface and the type of low energy excitations.}

In this subsection, we will turn to discuss the
case with $p<p_c$. We intend to  see the effect
of the dipole coupling on the Fermi surface as
well as its type, and investigate how the Fermi
surface enters into the oscillating region,
losing its meaning in the background with
hyperscaling violation. To study the solutions
explicitly, we will numerically solve the Dirac
equations to determine where the system possesses
a Fermi surface and the type of excitations.

\begin{figure}
\center{
\includegraphics[scale=0.5]{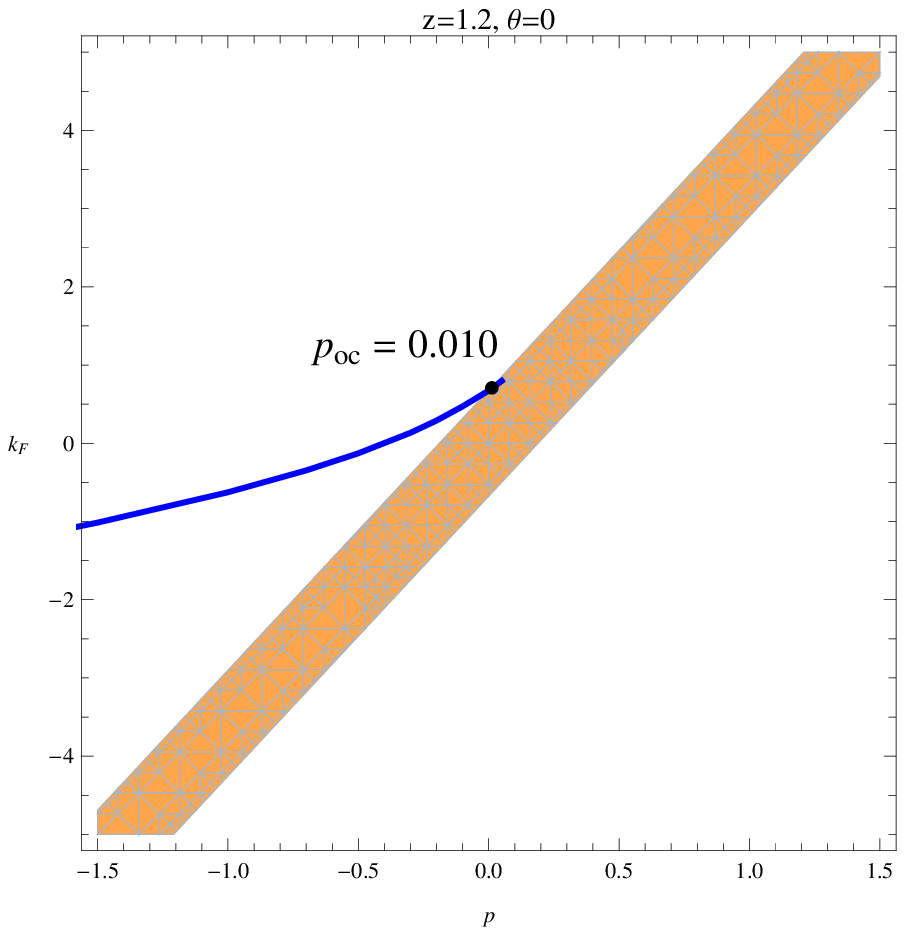}\hspace{0.4cm}
\includegraphics[scale=0.5]{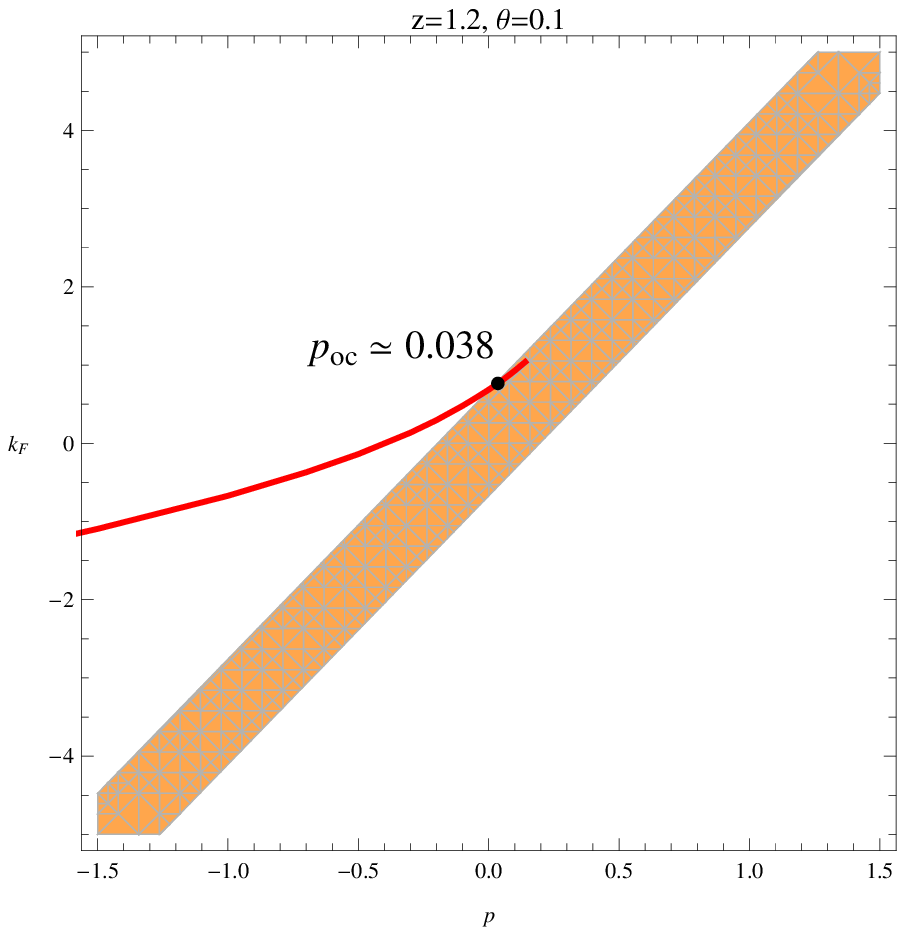}\hspace{0.4cm}
\includegraphics[scale=0.5]{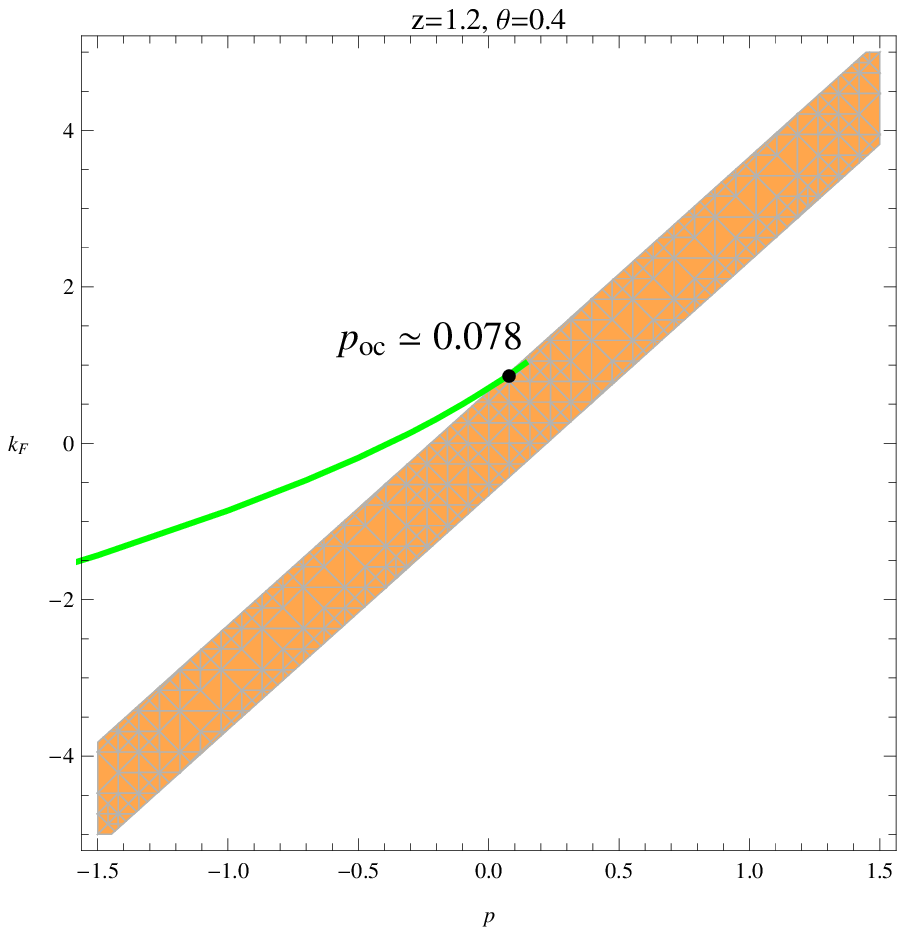}\hspace{0.4cm}
\caption{\label{figp-kF} The momentum of Fermi surface
corresponding to $G_2$ for  small dipole coupling. The orange band
denote the oscillatory region of $G_2$. The parameters from left
to right are $z=1.2$ and $\theta=0$, $z=1.2$ and $\theta=0.1$,
$z=1.2$ and $\theta=0.4$ respectively.}}
\end{figure}

Fig.~\ref{figp-kF} displays our findings for the
momentum of Fermi surface as well as its
behaviour entering into the oscillatory regions
with various exponents. In the figure we can see
that as the hyperscaling violation exponent
increases,  the critical dipole coupling $p_{oc}$
which makes the Fermi momentum to enter the
oscillatory region becomes larger due to the
smoother oscillating boundary. This behavior is
expected the same as that of $p_c$ to open the
Mott gap. The values of the critical dipole
couplings are $p_{oc}\simeq 0.010$ for $z=1.2$
and $\theta=0$, $p_{oc}\simeq 0.038$ for $z=1.2$
and $\theta=0.1$ while $p_{oc}\simeq 0.078$ for
$z=1.2$ and $\theta\simeq0.4$. In all cases, when
$p > p_{oc}$, the Fermi momentum will lose its
meaning of Fermi surface. When $p$ is smaller
than $p_{oc}$, the Fermi momentum increases as
the dipole coupling becomes large and then enters
into the oscillatory region, which is consistent
with the observation in \cite{Edalati:2010b}. In
addition, we show the samples of Fermi momentum
with different $p$ for various parameters in
Table.~\ref{tablep-kf}. For $p$ deviates away
from $p_{oc}$, we see that the Fermi momentum is
smaller for larger hyperscaling exponents. While
near the oscillating boundary and for the minimal
coupling, the Fermi momentum is larger for bigger
exponents due to the lower effective dimension as
we discussed in \cite{Kuang:2014}. This behavior
near the oscillatory region  can be attributed to
the smooth boundary of the region with larger
hyperscaling exponent.
\begin{center}\label{tablep-kf}
\begin{table}[ht]
\begin{tabular}{|c|c|c|c|c|c|c|c|}\hline
$p$ & $-1.5$ & $-1$ & $-0.5$ & $0$ & $0.02$ & $0.05$ & $0.1$   \\ \hline
$z=1.2, \theta=0$ &$-1.0141$ &$-0.6262$& $-0.1279$& $0.6812$& $-$& $-$& $-$ \\ \hline
$z=1.2, \theta=0.1$ &$-1.0942$ &$-0.6713$& $-0.1383$& $0.6883$& $0.7274$& $-$& $-$ \\ \hline
$z=1.2, \theta=0.4$ &-1.4334& $-0.8601$& $-0.1853$& $0.7084$ & $0.7490$ & $0.8120$& $-$ \\ \hline
\end{tabular}
\caption{\label{tablep-kf}  The Fermi momentum with different $p$
for the various parameters. Here $`-'$  denote the Fermi momentum
can not represent the Fermi surface.}
\end{table}
\end{center}

Having the Fermi momentum, we can calculate the
dimensionless scaling $\nu_2(k)$ in terms of the
expression (\ref{nuk}). As discussed in
\cite{Liu:2009}, imaginary $\nu_{I}(k=k_F)$
corresponds to ``log oscillatory'' solutions as
we emphasized before. When $\nu_{I}(k=k_F) <
1/2$, the pole of $G_R$ corresponds to an
unstable quasi-particle which is identified as a
non-Fermi fluid.  With the value $\nu_{I}(k=k_F)
=1/2$, the excitations are of marginal Fermi
fluid type. For $\nu_{I}(k=k_F) > 1/2$ the
dispersion relation is linear denoting the Fermi
fluid.
\begin{figure}
\center{
\includegraphics[scale=0.8]{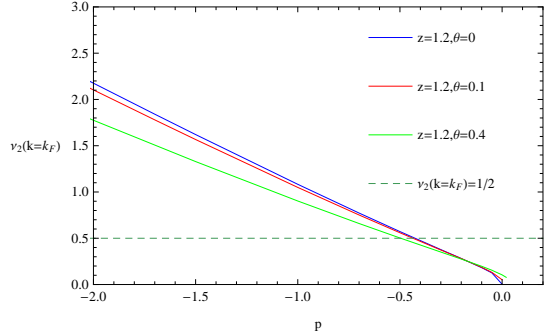}\hspace{0.5cm}
\caption{\label{fignuk-p} $\nu_2(k)$ changing with the dipole coupling for
 various parameter. We mark the marginal Fermi type with dashed line.}}
\end{figure}

The results of $\nu_2(k)$ changing with the
dipole coupling  are shown  in
Fig.~\ref{fignuk-p}. With all choices of
exponents, there is a phase transition from Fermi
liquid to marginal Fermi liquid then to non-Fermi
liquid as the dipole coupling becomes stronger.
Combining Fig.~\ref{fignuk-p} and
Fig.~\ref{figp-kF}, we can cconclude that the
system can not show a Fermi surface unless
$p<0.010$, thereafter the types of excitations
near the Fermi surface  are non-Fermi liquid for
$-0.425<p<0.010$, marginal Fermi liquid for
$p\simeq-0.425$ and Fermi liquid for $p<-0.425$
when $z=1.2$ and $\theta=0$. Taking the same $z$,
for example $z=1.2$, we find that different
hyperscaling values $\theta=0.1$ and $\theta=0.4$
lead to the values of marginal Fermi liquid
$p\simeq-0.441$ and $p\simeq-0.496$,
respectively. This tells us that smaller dipole
coupling corresponds to Fermi Liquid while larger
one corresponds to non-Fermi liquid ending by the
critical value $p_{oc}$.

With $z=1.2$, for larger hyperscaling exponent,
the phase transition happens at more negative
dipole coupling. We can understand the phenomenon
as follow.  As we mentioned earlier that larger
hyperscaling violation recedes the effective
dimension $d_{\theta}=2-\theta$, which always
appears in the form $p(z+d_{\theta}-2)$ in the
flow equation and compensate the effect of $p$
\cite{Kuang:2012}. So it is reasonable that lower
effective dimension corresponds to more negative
dipole coupling for the phase transition.
Combining the phenomena that larger $\theta$
needs more negative $p$ for the phase transition,
we observed that  the generation of gap with
larger $\theta$ calls for stronger positive $p_c$
in the last subsection. It would be interesting
to check the duality under the transformation
$p\rightarrow -p$ proposed in
\cite{Alsup:2014uca} for the hyperscaling model
which will be carried out in the next section.

\section{The zeros-poles duality of the holographic system}\label{SDuality}
In this section we will study  the behaviour of $\rm{det~G_R}$ in
different dipole couplings with the aim to see if in holographic
theories with hyperscaling violation the duality between zeros and
poles under $p\rightarrow -p$ found in \cite{Alsup:2014uca} still
holds.   Following \cite{Alsup:2014uca} we define
\begin{equation}
\zeta_{I}=\frac{1}{\xi_{I}}
\end{equation}
which satisfies the following equation
\begin{eqnarray} \label{floweq2}
\left(\sqrt{f}\partial_{u}+2m u^{\frac{\theta}{2}-1}\right) \zeta_{I}
-\left[ \frac{\tilde{v_{-}}}{u} + (-1)^{I} k  \right]\zeta_{I}^{2}
- \left[ \frac{\tilde{v_{+}}}{u} - (-1)^{I} k  \right]
=0~.
\end{eqnarray}
Comparing the above equation with equation
(\ref{DiracEF2}), we find that the equation of
$\zeta_{I}$ coincides with the equation of
$\xi_{I}$ under the transformation of
$(m,k,p)\rightarrow(-m,-k,-p)$. Then considering
the symmetry of the Green function (\ref{Gsym}),
we can get the relation
\begin{eqnarray} \label{duality}
\rm{det}G_R(\omega=0,k;m,p)=\rm{det}G_R(\omega=0,-k;m,p)=\frac{1}{\rm{det}G_R(\omega=0,k;-m,-p)}~.
\end{eqnarray}
The formula with $m=0$ coincides with the expression (30) in
\cite{Alsup:2014uca} and its value is one  for $p=0$
\cite{Faulkner:2009}.

Note that a pole of $G_I$ at $\omega =0$  is not necessarily a
pole of the determinant $ \rm{det} G_R = G_{1} G_{2}. $ It is
known \cite{Faulkner:2010da} that in the conventional case, $p=0$,
$ \rm{det} G_R (\omega =0 , k; p=0) =-1, $ therefore it possesses
neither poles nor zeroes. This is because poles (zeroes) of $G_1$
are cancelled by zeroes (poles) of $G_2$ at the same momentum. It
was showed in \cite{Alsup:2014uca} that this coincidence of poles
and zeroes is lifted when the dipole coupling is turned on,
resulting in poles and zeroes of $\det G_R$.

\begin{figure}
\center{
\includegraphics[scale=0.5]{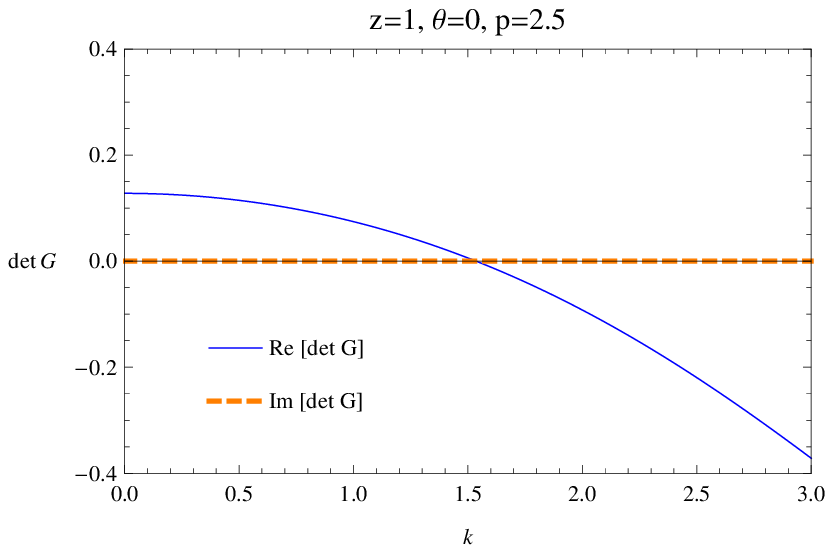}
\includegraphics[scale=0.5]{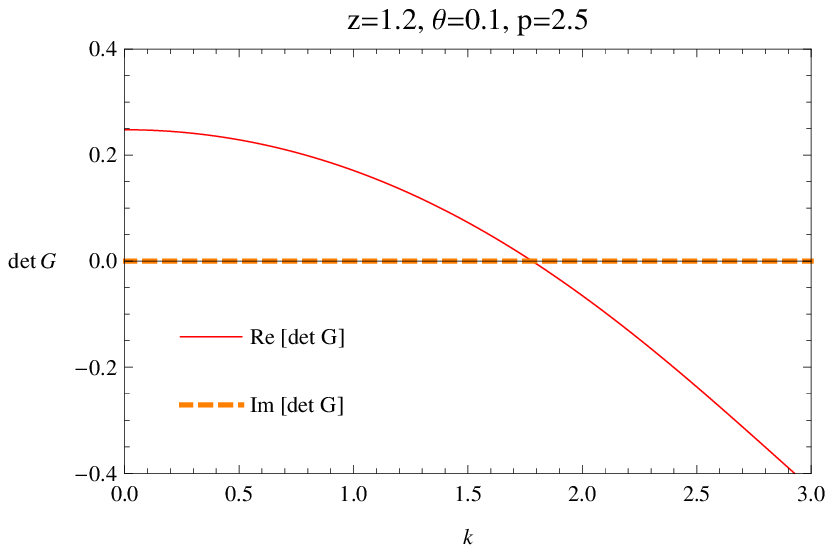}
\includegraphics[scale=0.5]{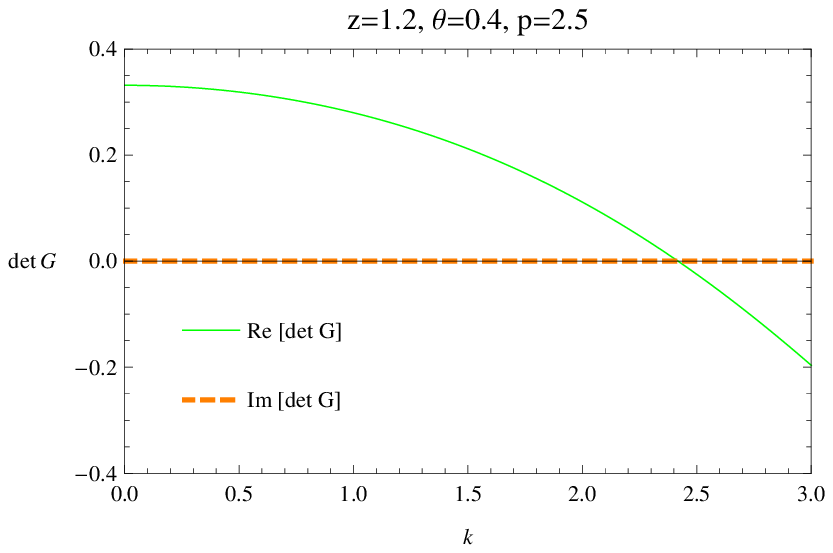}
\includegraphics[scale=0.5]{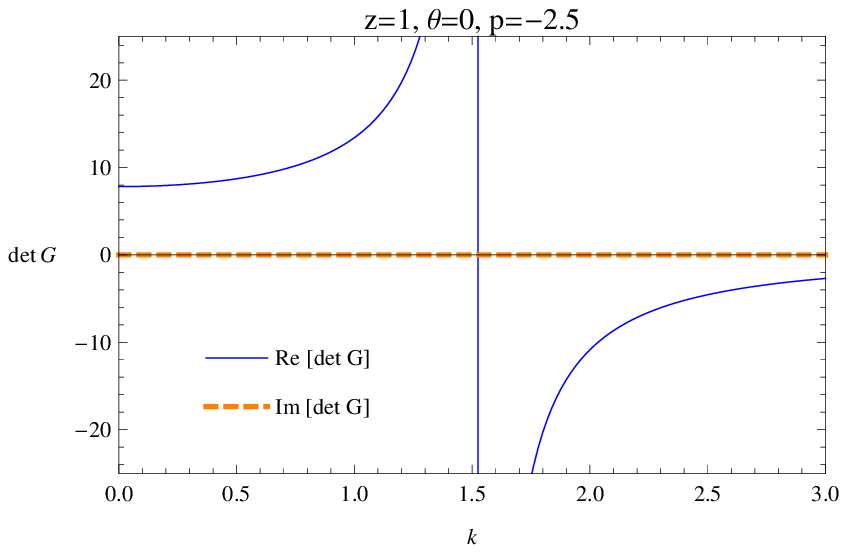}
\includegraphics[scale=0.5]{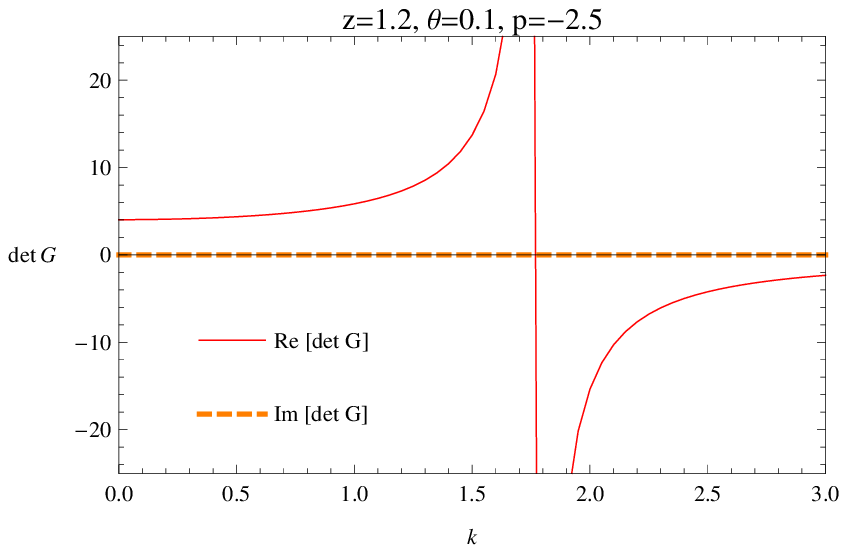}
\includegraphics[scale=0.5]{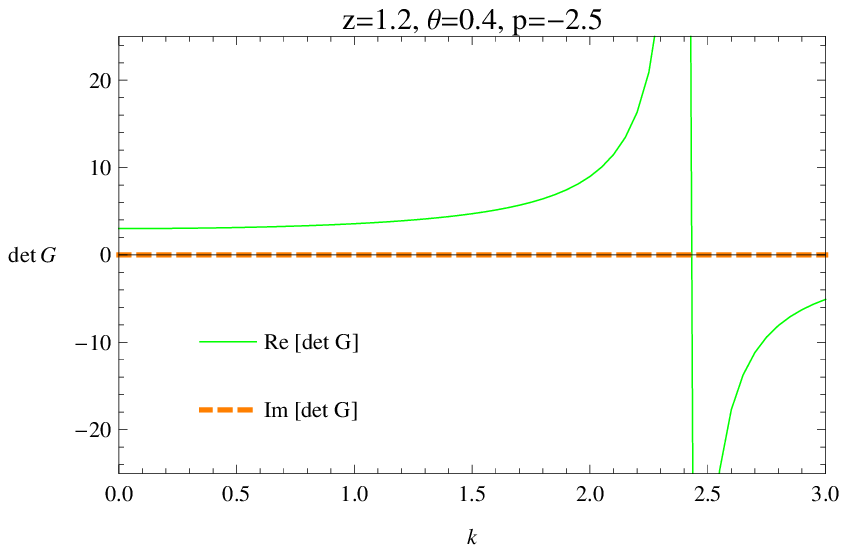}
\caption{\label{figk-detG} The real and imaginary parts of
$\rm{det G}$ depending on the momentum for $p=2.5$ with
hyperscaling violation.}}
\end{figure}

We firstly turn off the hyperscaling violation factor. The result
is showed in the left panel of Fig.~\ref{figk-detG} with $p=2.5$,
which we reproduce the corresponding result  of
\cite{Alsup:2014uca} with $p=5$. The behaviour of $\rm{det G}$
with non-zero hyperscaling violation is also presented in the
middle and right panel of Fig.~\ref{figk-detG}. For $\theta=0.1$,
the zero for the real part of $\rm{det G}$ is around $k\simeq
1.78$ with $p=2.5$, corresponding to a pole at the same momentum
with $p=-2.5$. While for $\theta=0.4$, the momentum related to the
zero-pole duality is $k\simeq 2.42$. We see that the hyperscaling
violation does not break the zero-pole duality under the
transformation $p\rightarrow -p$. This is expected because the
zeros-poles duality is a reflection of the symmetries of the
Green's function and it should not be related to the
dimensionality of these theories.

\section{Conclusions and discussion}\label{SConclusion}

We have studied the behaviour of a holographic fermionic system
 with a charged black brane with
hyperscaling violation in the bulk in the
presence of dipole interaction between a massless
fermion and a gauge field. We found that the
holographic system possesses a robust phase
diagram including Fermi and non-Fermi liquids,
marginal Fermi liquid, log oscillatory phase and
an  insulating Mott phase.  These various phases
are controlled by the strength of the dipole
coupling and the hyperscaling violation exponent
which
  play the role of the order parameters in the
holographic system \cite{Edalati:2010a,Edalati:2010b}.

These phases can be identified by studying   the
behaviour of the Green's function. Considering
the IR and the UV limits of the Green's function,
we used the matching method \cite{Iqbal:2009} in
the  near horizon geometry to obtain the
analytical expressions of the UV Green's function
and the dispersion relation.

We found that as   the hyperscaling violation
exponent is increased the critical value of the
dipole moment $p_c$ for a Mott gap to be
generated is also increased, which  makes it
harder for the insulating phase to be formed.
This is attributed to the fact that a larger
hyperscaling violation exponent corresponds to a
lower effective dimensionality of the system and
this change compensates the increase in critical
dipole coupling $p_{c}$. Then, we investigated
the existence of the Fermi surface and the
disperse relation near the Fermi surface. We
found that as $p$ increases, the liquid changes
from Fermi type, marginal Fermi type to non-Fermi
type, then the Fermi surface disappears in the
Log-oscillatory region. The phase transition
between  the types of liquid happens at negative
dipole coupling and the larger hyperscaling
violation will result in more negative dipole
coupling.

A pole is indicative of a (non-) Fermi fluid while a zero is
responsible for an insulating phase. It is the coexistence of both
that underlies the various phases of the liquid. We showed that
the  duality found in \cite{Alsup:2014uca} relating systems of
opposite dipole coupling strength $p$ also persists in holographic
phase with hyperscaling violation exponent.

It would be interesting to consider a Lorentz violating boundary
term into the bulk action instead of the standard boundary
condition. It was discussed in \cite{Laia:2011} that with the
Lorentz violating boundary term, the dual field theory has a
holographic non-relativistic fixed point, possessing a flat band
of gapless excitation. With the minimal coupling, we have also
observed a flat band  in the boundary theory dual to hyperscaling
violation background in \cite{Kuang:2014}. So it is important to
see how the hyperscaling violation will influence  the dipole
effect in the dual non-relativistic fermionic system. This study
is underway.

Another possible direction is to calculate the
holographic entanglement entropy in our theory.
The entanglement entropy
\cite{Ryu:2006bv,Ryu:2006ef} was proved to be a
powerful tool in counting the degrees of freedom
available in a holographic system. In
\cite{Kuang:2014kha} it was found that the
holographic entanglement entropy in the
superconducting phase is less than that in the
normal phase due to the fact that Cooper pairs
had been formed so that fewer degrees of freedom
are aleft(see also \cite{Albash:2012pd}). Near
the contact interface of the superconductor to
normal metal,  the entanglement entropy could
have  higher value in the superconducting phase
due to the proximity effect: the leakage of
Cooper pairs to the normal phase results in more
freedom in the superconductor side near the
interface. It would be interesting to generalize
previous study and discuss the holographic
entanglement entropy in a holographic fermionic
system with a dipole coupling and hyperscaling
violation. This can give important information on
the various phases of the system. To carry out
such a study, we need a fully back-reacted
solution of the Einsten-Maxwell-Dirac system. We
will report results on this topic in the future.

\begin{acknowledgments}
X. M. K. is indebted to Li-Qing Fang for
helping to run parts of our Mathematica programs in his computer. X. M. K. and E. P. are supported by
ARISTEIA II action of the operational programme
education and long life learning which is
co-funded by the European Union (European Social
Fund) and National Resources. B. W. and J. P. W. are
supported by the Natural Science Foundation of China.
J. P. W. is also supported by Program for Liaoning Excellent Talents in University (No. LJQ2014123).\\
\end{acknowledgments}


\begin{thebibliography}{99}
\bibitem{Maldacena:1997re}
J.~M.~Maldacena, \emph{The large N limit of superconformal field
theories and supergravity}, Adv.\ Theor.\ Math.\ Phys.\  {\bf 2}
(1998) 231 [Int.\ J.\ Theor.\ Phys.\  {\bf 38} (1999) 1113].

\bibitem{Gubser:2002}
 S. S. Gubser, I. R. Klebanov and A. M.
Polyakov, \emph{A semiclassical limit of the gauge string
correspondence}, Nucl. Phys. B\textbf{ 636} (2002) 99.

\bibitem{Witten:1998} E. Witten, \emph{Anti-de Sitter space and
holography}, Adv. Theor. Math. Phys. \textbf{2} (1998) 253.


\bibitem{Lee:2008}
S. S. Lee, \emph{A non-Fermi liquid from a charged black hole: a critical Fermi ball},
Phys. Rev. D\textbf{ 79} (2009) 086006, [arXiv:0809.3402].

\bibitem{Liu:2009}
H. Liu, J. McGreevy and D. Vegh, \emph{Non-Fermi liquids from holography},
Phys. Rev. D \textbf{ 83} (2011) 065029 [arXiv:0903.2477].

\bibitem{Cubrovic:2009ye}
M.~Cubrovic, J.~Zaanen, and K.~Schalm, ``{String Theory, Quantum
Phase
  Transitions and the Emergent Fermi-Liquid}, '' {\em Science} {\bf 325} (2009)
  439--444, \href{http://xxx.lanl.gov/abs/0904.1993}{{\tt 0904.1993}}.

\bibitem{Faulkner:2009}
T. Faulkner, H. Liu, J. McGreevy and D. Vegh, \emph{Emergent quantum criticality, Fermi
surfaces and $AdS_{2}$}, Phys. Rev. D\textbf{ 83} (2011) 125002, [arXiv:0907.2694].

\bibitem{Edalati:2010a}
M. Edalati, R.G. Leigh and P.W. Phillips, \emph{Dynamically generated Mott gap from holography},
Phys. Rev. Lett. \textbf{106} (2011) 091602, [arXiv:1010.3238].

\bibitem{Edalati:2010b}
M. Edalati, R.G. Leigh, K.W. Lo and P.W. Phillips, \emph{Dynamical gap and cuprate-like physics
from holography}, Phys. Rev. D\textbf{ 83} (2011) 046012, [arXiv:1012.3751].

\bibitem{Guarrera:2011my}
  D.~Guarrera and J.~McGreevy,
  \emph{Holographic Fermi surfaces and bulk dipole couplings},
  [arXiv:1102.3908].

\bibitem{Wu:2011a}
J. P. Wu, \emph{Holographic fermions in charged Gauss-Bonnet black
hole}, JHEP \textbf{07} (2011) 106 [arXiv:1103.3982];X. M. Kuang, B.
Wang, J. P. Wu,  \emph{Dynamical gap from holography in the
charged dilaton black hole}, Class. Quant. Grav. \textbf{30}
(2013) 145011, [arXiv:1210.5735].
\bibitem{Kuang:2012}
X. M. Kuang, B. Wang, J. P. Wu, \emph{Dipole coupling effect of holographic
fermion in the background of charged Gauss-Bonnet AdS black hole},
JHEP \textbf{07} (2012) 125, [arXiv:1205.6674].


\bibitem{Fang:2013}
L. Q. Fang, X. H. Ge, X. M. Kuang, \emph{Holographic fermions with
running chemical potential and dipole coupling}, Nucl. Phys. B
\textbf{ 877} (2013) 807-824, [arXiv:1304.7431].

\bibitem{Wu:2011b}
J. P. Wu, \emph{Some properties of the holographic fermions in an
extremal charged dilatonic black hole}, Phys. Rev. D\textbf{ 84}
064008, (2011), [arXiv:1108.6134]; W. J. Li, J. P. Wu,
\emph{Holographic fermions in charged dilaton black branes},
Nuclear Physics B \textbf{ 867} (2013), 810-826 [arXiv:1203.0674];
  W.~J.~Li, R.~Meyer and H.~b.~Zhang,
  \emph{Holographic non-relativistic fermionic fixed point by the charged dilatonic black hole},
  JHEP {\bf 1201}, 153 (2012)
  [arXiv:1111.3783].
\bibitem{Li:2011}
W. J. Li, H. Zhang, \emph{Holographic non-relativistic fermionic
fixed point and bulk dipole coupling}, JHEP, \textbf{1111}, 018
(2011), [arXiv:1110.4559].
\bibitem{Ling:2014}
Y. Ling, P. Liu, C. Niu, J.~P. Wu, Z. Y. Xian, \emph{Holographic fermionic system with dipole coupling on Q-lattice}, [arXiv:1410.7323].
\bibitem{Alsup:2014uca}
  J.~Alsup, E.~Papantonopoulos, G.~Siopsis and K.~Yeter,
  \emph{Duality between zeroes and poles in holographic systems with massless fermions and a dipole coupling},
  [arXiv:1404.4010].

\bibitem{Vanacore:2014hka}
  G.~Vanacore and P.~W.~Phillips,
  \emph{Minding the Gap in Holographic Models of Interacting Fermions},
  [arXiv:1405.1041].

\bibitem{Kachru:2008} S. Kachru, X. Liu and M. Mulligan, \emph{Gravity Duals of Lifshitz-like Fixed Points,} Phys. Rev. D 78
(2008) 106005, [arXiv:0808.1725].
\bibitem{Gouteraux:2011} B. Gouteraux and E. Kiritsis, \emph{Generalized Holographic Quantum Criticality at Finite Density,} JHEP
1112 (2011) 036, [arXiv:1107.2116].
\bibitem{Huijse:2012} L. Huijse, S. Sachdev and B. Swingle, \emph{Hidden Fermi surfaces in compressible states of gauge-gravity
duality,} Phys. Rev. B 85, 035121 (2012), [arXiv:1112.0573].
\bibitem{Dong:2012}X. Dong, S. Harrison, S. Kachru, G. Torroba and H. Wang, \emph{Aspects of holography for theories with
hyperscaling violation,} JHEP 1206 (2012) 041, [arXiv:1201.1905].

\bibitem{Charmousis:2010}
C. Charmousis, B. Gouteraux, B. S. Kim, E. Kiritsis and R. Meyer, \emph{Effective Holographic Theories
for low-temperature condensed matter systems}, JHEP 1011, 151 (2010) [arXiv:1005.4690].
\bibitem{Fisher}D. S. Fisher, \emph{Scaling and critical slowing down in random-field Ising systems}, Phys.
Rev. Lett. 56, 416 (1986).







\bibitem{Taylor:2008}
  M.~Taylor,
  \emph{Non-relativistic holography},
  [arXiv:0812.0530].
\bibitem{Pang:2009ad}
  D.~W.~Pang,
  \emph{A Note on Black Holes in Asymptotically Lifshitz Spacetime},
  Commun.\ Theor.\ Phys.\  {\bf 62}, 265 (2014)
  [arXiv:0905.2678].

\bibitem{Tarrio:2011}
J. Tarrio, S. Vandoren, \emph{Black holes and black branes in Lifshitz spacetimes}, JHEP \textbf{1109} (2011) 017, [arXiv:1105.6335].

\bibitem{Ogawa:2011}N. Ogawa, T. Takayanagi and T. Ugajin, \emph{Holographic Fermi Surfaces and Entanglement Entropy,}
JHEP 1201, 125 (2012), [arXiv:1111.1023].

\bibitem{Wolf:2005} M. M. Wolf, \emph{Violation of the entropic area law for Fermions,} Phys. Rev. Lett. 96, 010404 (2006),
[quant- ph/0503219]; B. Swingle, \emph{Entanglement Entropy and
the Fermi Surface,} Phys. Rev. Lett. 105, 050502 (2010), [arXiv:0908.1724].

\bibitem{Huijse:2011ef}
  L.~Huijse, S.~Sachdev and B.~Swingle,
  ``Hidden Fermi surfaces in compressible states of gauge-gravity duality,''
  Phys.\ Rev.\ B {\bf 85}, 035121 (2012)
  [arXiv:1112.0573 [cond-mat.str-el]].

\bibitem{Gursoy:2011}
U. Gursoy, E. Plauschinn, H. Stoof, S. Vandoren, \emph{Holography and ARPES
sum-rules}, JHEP \textbf{1205} (2012) 018, [arXiv:1112.5074].

\bibitem{Alishahiha:201201}
  M.~Alishahiha, M.~R.~Mohammadi Mozaffar and A.~Mollabashi,
  \emph{Fermions on Lifshitz Background},
  Phys.\ Rev.\ D {\bf 86}, 026002 (2012)
  [arXiv:1201.1764].
\bibitem{Alishahiha:201209}
Mohsen Alishahiha, Eoins O Colgsain, Hossein Yavartanoo,
\emph{Charged Black Branes with Hyperscaling Violating Factor},
JHEP \textbf{11}(2012)137,[arXiv:1209.3946].

\bibitem{Fang:2012}
L. Q. Fang, X. H. Ge and X. M. Kuang, \emph{Holographic fermions in charged Lifshitz theory}, Phys. Rev. D \textbf{86}, 105037 (2012)
[arXiv:1201.3832].

\bibitem{Kuang:2014}
X.-M. Kuang, E. Papantonopoulos, B. Wang, J.-P. Wu, \emph{Formation of Fermi surfaces and the appearance of liquid phases in holographic theories with hyperscaling violation,} [arXiv:1409.2945].


\bibitem{wu-lifshitz}
J. P. Wu, \emph{Holographic fermions on a charged Lifshitz background from Einstein-Dilaton-Maxwell model}, JHEP \textbf{03} (2013) 083; J. P. Wu, \emph{Emergence of gap from holographic fermions on charged Lifshitz background}, JHEP \textbf{04} (2013) 073; J. P. Wu, \emph{The charged Lifshitz black brane geometry and the bulk dipole coupling}, Phys. Lett. B \textbf{728} (2014) 450-456.




%
%



%
%


\bibitem{Faulkner:2010da}
  T.~Faulkner, N.~Iqbal, H.~Liu, J.~McGreevy and D.~Vegh,
  ``From Black Holes to Strange Metals,''
  arXiv:1003.1728 [hep-th].

\bibitem{Iqbal:2009}
N. Iqbal, H. Liu, \emph{Real-time response in AdS/CFT with application to spinors},  Fortsch. Phys. 57:367-384, 2009, [arXiv:0903.2596].

\bibitem{Laia:2011}
  J.~N.~Laia and D.~Tong,
  ``A Holographic Flat Band,''
  JHEP {\bf 1111}, 125 (2011)
  [arXiv:1108.1381 [hep-th]].




\bibitem{Ryu:2006bv}
  S.~Ryu and T.~Takayanagi,
  ``Holographic derivation of entanglement entropy from AdS/CFT,''
  Phys.\ Rev.\ Lett.\  {\bf 96}, 181602 (2006)
  [hep-th/0603001].

\bibitem{Ryu:2006ef}
  S.~Ryu and T.~Takayanagi,
  ``Aspects of Holographic Entanglement Entropy,''
  JHEP {\bf 0608}, 045 (2006)
  [hep-th/0605073].

\bibitem{Kuang:2014kha}
  X.~-M.~Kuang, E.~Papantonopoulos and B.~Wang,
  ``Entanglement Entropy as a Probe of the Proximity Effect in Holographic Superconductors,''
  JHEP {\bf 1405}, 130 (2014)
  [arXiv:1401.5720 [hep-th]].

\bibitem{Albash:2012pd}
  T.~Albash and C.~V.~Johnson,
 ``Holographic Studies of Entanglement Entropy in Superconductors,''
  JHEP {\bf 1205}, 079 (2012)
  [arXiv:1202.2605 [hep-th]].


\bibitem{0912.1061}
S.A. Hartnoll, J. Polchinski, E. Silverstein and D. Tong, Towards strange metallic
holography, JHEP 04 (2010) 120 [arXiv:0912.1061].

\end{thebibliography}
\end{document}